\documentclass[reprint,superscriptaddress,nofootinbib,twocolumn,amsmath,amssymb,aps,prl]{revtex4-2}

\newcommand{\bra}[1]{\langle#1|}
\newcommand{\ket}[1]{|#1\rangle}
\newcommand{\braket}[1]{\langle#1\rangle}

\DeclareMathOperator{\sinc}{sinc}

\usepackage{graphicx}
\usepackage{dcolumn}
\usepackage{float}
\usepackage{bm}
\usepackage[colorlinks]{hyperref}
\usepackage[utf8]{inputenc}
\usepackage{lmodern} 
\usepackage{siunitx}
\usepackage[version=4]{mhchem}
\usepackage{microtype}
\usepackage{gensymb}

\begin{document}

\preprint{APS/123-QED}

\title{Emergence of multiphoton quantum coherence by light propagation}

\author{Jannatul Ferdous}
\affiliation{Quantum Photonics Laboratory, Department of Physics \& Astronomy, Louisiana State University, Baton Rouge, LA 70803, USA}

\author{Mingyuan Hong}
\affiliation{Quantum Photonics Laboratory, Department of Physics \& Astronomy, Louisiana State University, Baton Rouge, LA 70803, USA}

\author{Riley B. Dawkins}
\affiliation{Quantum Photonics Laboratory, Department of Physics \& Astronomy, Louisiana State University, Baton Rouge, LA 70803, USA}

\author{Fatemeh Mostafavi}
\affiliation{Quantum Photonics Laboratory, Department of Physics \& Astronomy, Louisiana State University, Baton Rouge, LA 70803, USA}

\author{Alina Oktyabrskaya}
\affiliation{Department of Mathematics, Louisiana State University, Baton Rouge, LA 70803, USA}

\author{Chenglong You}
\email{cyou2@lsu.edu}
\affiliation{Quantum Photonics Laboratory, Department of Physics \& Astronomy, Louisiana State University, Baton Rouge, LA 70803, USA}

\author{Roberto de J. Le\'on-Montiel}
\affiliation{Instituto de Ciencias Nucleares, Universidad Nacional Aut\'onoma de M\'exico, Apartado Postal 70-543, 04510 Cd. Mx., M\'exico}

\author{Omar S. Maga\~na-Loaiza}

\affiliation{Quantum Photonics Laboratory, Department of Physics \& Astronomy, Louisiana State University, Baton Rouge, LA 70803, USA}

\date{\today}

\begin{abstract}
The modification of the quantum properties of coherence of photons through their interaction with matter lies at the heart of the quantum theory of light. Indeed, the absorption and emission of photons by atoms can lead to different kinds of light with characteristic quantum statistical properties. As such, different types of light are typically associated with distinct sources. Here, we report on the observation of the modification of quantum coherence of multiphoton systems in free space. This surprising effect is produced by the scattering of thermal multiphoton wavepackets upon propagation. The modification of the excitation mode of a photonic system and its associated quantum fluctuations result in the formation of different light fields with distinct quantum coherence properties. Remarkably, we show that these processes of scattering can lead to multiphoton systems with sub-shot-noise quantum properties. Our observations are validated through the nonclassical formulation of the emblematic van Cittert-Zernike theorem. We believe that the possibility of producing quantum systems with modified properties of coherence, through linear propagation, can have dramatic implications for diverse quantum technologies.
\end{abstract}

\maketitle
\section{Introduction}
The description of the evolution of spatial, temporal, and polarization properties of the light field gave birth to the classical theory of optical coherence \cite{wolf1954optics, mandel1995optical, born2013principles, Dorrer:04, gori2000use, Cai2020, Cai2012}. Naturally, these properties of light can be fully characterized through the classical electromagnetic theory \cite{born2013principles}. Furthermore, there has been interest in describing the evolution of propagating quantum optical fields endowed with these classical properties \cite{Saleh05PRL, you2023multiphoton}. This has been accomplished by virtue of the Wolf equation and the van Cittert-Zernike theorem \cite{wolf1954optics, Saleh05PRL, Cittert:34, ZERNIKE:38}. Nevertheless, there is a long-sought goal in describing the evolution of the inherent quantum mechanical properties of the light field that define its nature and kind \cite{GlauberPR1963, PhysRevLettSudarshan}. Such formalism would enable modeling the evolution of the excitation mode of propagating electromagnetic fields in the Fock number basis. Given the large number of scattering and interference processes that can take place in a quantum optical system with many photons, this ambitious description remains elusive \cite{anno2006, magana2019multiphoton, You2020plasmonics, you2021observation, Walmsley:23}. Although, it is essential to describe the evolution of propagating multiphoton wavepackets in diverse quantum photonic devices \cite{Walmsley:23, Omar2019Review, bhusal2021smart, aspuru-guzik_photonic_2012}. 

\medskip
\medskip

The quantum theory of optical coherence developed by Glauber and Sudarshan provides a description of the excitation mode of the electromagnetic field \cite{PhysRevLettSudarshan, GlauberPR1963, PhysRevCoherentIncoherent}. This elegant formalism led to the identification of different kinds of light that are characterized by distinct quantum statistical fluctuations and noise levels \cite{magana2019multiphoton, you2020identification, GlauberPR1963, PhysRevCoherentIncoherent,YouLXAP2024}.  As such, a particular quantum state of light is typically associated with a specific emission process and a light source \cite{PhysRevCoherentIncoherent}. Moreover, the quantum theory of electromagnetic radiation enables describing light-matter interactions \cite{allen1987optical}. These consist of absorption and emission processes that can lead to the modification of the excitation mode of the light field and consequently to different kinds of light \cite{allen1987optical, PhysRevCoherentIncoherent}. This possibility has triggered interest in achieving optical non-linearities at the single-photon level to engineer and control quantum states of light \cite{venkataraman_phase_2013, Zasedatelev-2021, Snijders-2016, HallajiNatPhys2017}. Thus, it is believed that the excitation mode of the light field, and its quantum properties of coherence, remain unchanged upon propagation in free space \cite{PhysRevCoherentIncoherent,allen1987optical}.

\begin{figure*}[!ht]
   \centering 
   \includegraphics[width=1.0\textwidth]{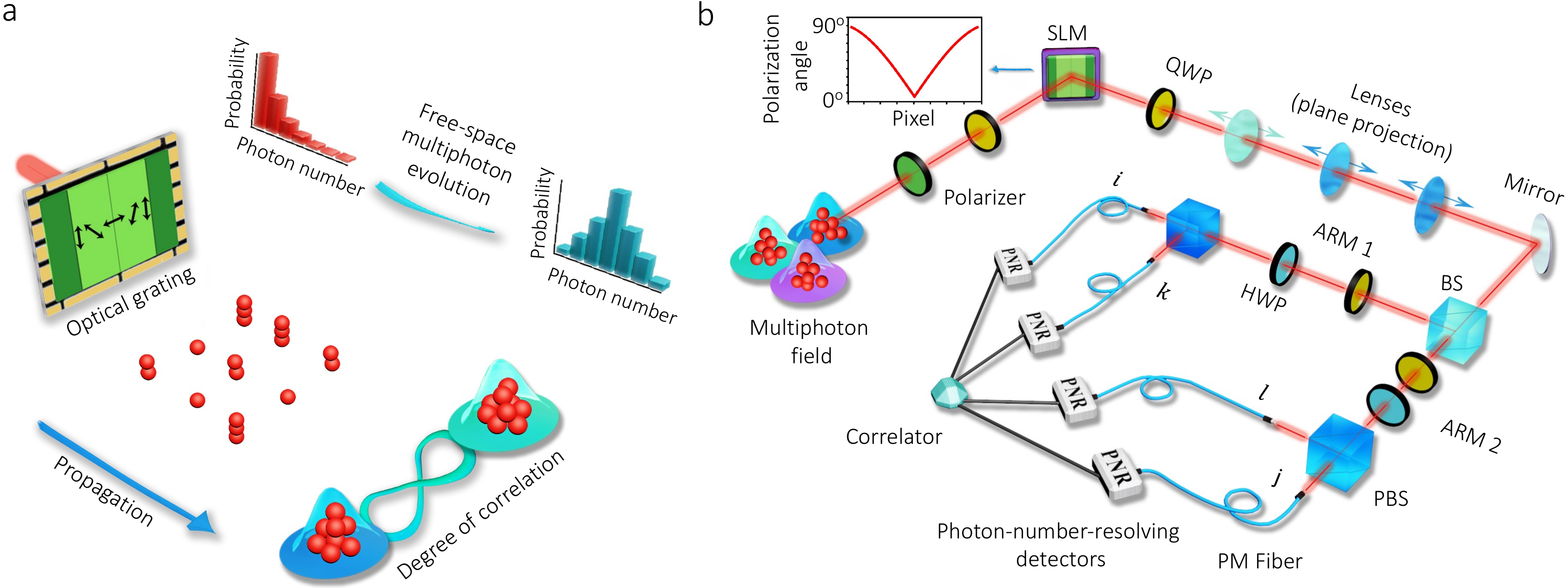}
    \caption{\textbf{Modification of quantum coherence in propagating multiphoton wavepackets.} 
    The diagram in \textbf{a} illustrates the scattering of
    thermal multiphoton wavepackets by an optical grating. The grating modifies the polarization properties of the multiphoton wavepackets at different transverse spatial locations. The interference of the scattered multiphoton wavepackets, at different propagation planes, leads to changes in the quantum statistical properties of the thermal field. Interestingly, these interference events lead to the modification of multiphoton quantum coherence upon propagation in free space. The setup for the experimental investigation of this effect is shown in \textbf{b}. Here, a multimode thermal multiphoton beam passes through a polarizer and a quarter-wave plate (QWP) to modulate its polarization. The transmitted circularly polarized photons illuminate a spatial light modulator (SLM) where we display a polarization grating. The beam reflected off the SLM is sent to another QWP to rotate its polarization at different transverse positions (details can be found in the SI) \cite{PRLMirhossein2014}. The resulting polarization angle as a function of the transverse pixel position is depicted next to the SLM. This experimental arrangement induces partial polarization properties to the initial thermal light beam. The multiphoton field is then sent to a tunable telescope consisting of three lenses. This setup enables us to select different propagation planes of the scattered multiphoton field. We then perform polarization tomography of multiphoton wavepackets at an specific propagation plane by means of a beam splitter (BS), half-wave plates (HWPs), QWPs, and two polarizing beam splitters (PBS) \cite{Altepeter2005PhotonicST}. We use photon-number-resolving (PNR) detection to characterize the quantum coherence of propagating multiphoton systems and their quantum fluctuations \cite{you2020identification, HashemiRafsanjani2017}. }
   \label{Fig. 1}
\end{figure*} 

We demonstrate that the statistical fluctuations of thermal light fields, and their quantum properties of coherence, can be modified upon propagation in the absence of light-matter interactions \cite{you2023multiphoton,Hong_Mingyuan}. This effect results from the scattering of multiphoton wavepackets that propagate in free space. The large number of interference effects hosted by multiphoton systems with up to twenty particles leads to a modified light field with evolving quantum statistical properties \cite{You2020plasmonics, anno2006}. Further, we show that the evolution of multiphoton quantum coherence can be described by the nonclassical formulation of the van Cittert-Zernike theorem \cite{you2023multiphoton}. Interestingly, our description of propagating multiphoton quantum coherence unveils conditions under which multiphoton systems with sub-shot-noise quantum properties are formed \cite{gerry2005introductory}. Remarkably, these quantum multiphoton systems are produced upon propagation in the absence of optical nonlinearities \cite{venkataraman_phase_2013, Zasedatelev-2021, Snijders-2016, HallajiNatPhys2017}. As such, we believe that our findings provide an all-optical alternative for the preparation of multiphoton systems with nonclassical statistics. Given the relevance of photonic quantum control for multiple quantum technologies, similar functionalities have been explored in nonlinear optical systems, photonic lattices, plasmonic systems, and Bose-Einstein condensates \cite {OLSEN2002373, you2021observation, kondakci2015, venkataraman_phase_2013, Zasedatelev-2021, Snijders-2016, HallajiNatPhys2017}.

\section{Experiment and Results Discussion}

The optical system under consideration is depicted in Figure \ref{Fig. 1}\textbf{a}. In this case, an unpolarized thermal field is scattered by an optical grating to produce multiphoton wavepackets with distinct polarization properties at different transverse spatial positions \cite{gori2000use}. The scattered photons contained in the thermal beam interfere upon propagation to change the statistical fluctuations of the field \cite{you2023multiphoton}. Interestingly, these effects enable the modification of the quantum properties of coherence of the initial multiphoton thermal system in free space. As discussed in the Supplementary Information (SI), we describe our initial thermal system as an incoherent superposition of coherent states \cite{PhysRevLettSudarshan, GlauberPR1963, ou2007multi} 

\begin{equation}
\label{eq:1}
    \hat{\rho} = \int d\Sigma \bigotimes_{\boldsymbol{s}} \Big(|\alpha\rangle\langle\alpha|_{\Sigma,H,\boldsymbol{s}}+|\alpha\rangle\langle\alpha|_{\Sigma,V,\boldsymbol{s}}\Big),
\end{equation}
where $|\alpha\rangle_{\Sigma,B,\boldsymbol{s}}$ represents a coherent state of amplitude $\alpha$ with random mode-structure $\Sigma$, where $\hat{a}_{\Sigma,B,\boldsymbol{s}} = \int d \boldsymbol{\rho} \text{Rect}[(\boldsymbol{s}-\boldsymbol{\rho})/d]\Sigma(\boldsymbol{\rho})\hat{a}_B(\boldsymbol{\rho})$ and polarization $B\in\{H,V\}$ (see SI). Furthermore, the tensor product over positions $\boldsymbol{s}$ represents the pixelated transverse beam profile where each pixel has sidelength $d$.

\begin{figure*}[t!]
   \centering 
   \includegraphics[width=\textwidth]{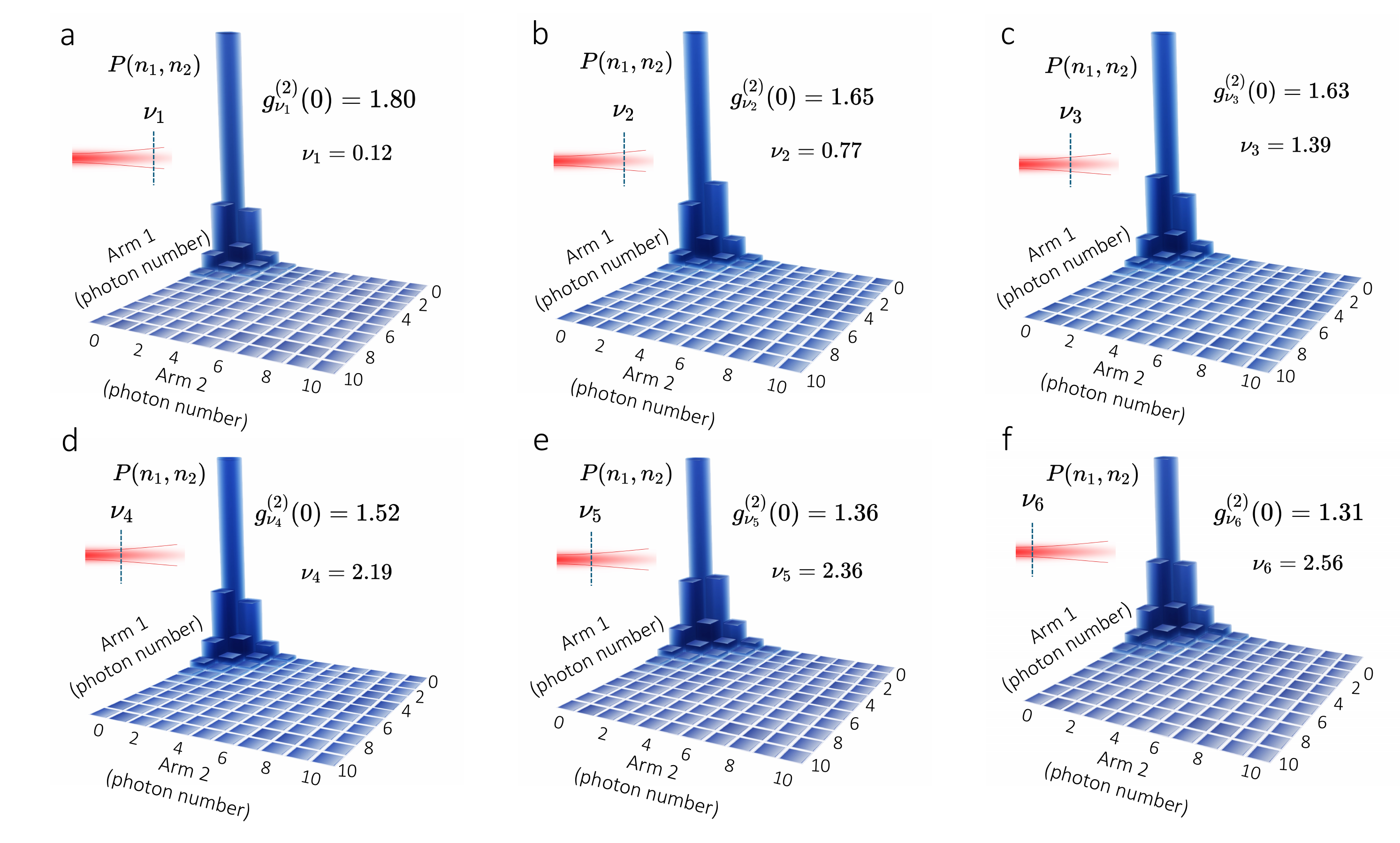}
    \caption{\textbf{Evolving quantum coherence induced by light propagation.} 
    The propagation of the multiphoton system reflected off the SLM induces modifications in its photon-number distribution. In this case, we focus on the horizontally-polarized component of the initial thermal beam with up to twenty particles. As shown in \textbf{a}, the multiphoton system at the propagation plane characterized by $\nu_{1} = 0.12$ is nearly thermal. We define $\nu$ as $L\Delta X/(\lambda z)$, in this case $L=3 \text{mm}$, $\Delta X=2 \text{mm}$, $\lambda=780 \text{nm}$ and we scan the propagation distance $z$.  Interestingly, the quantum fluctuations of the multiphoton system are attenuated with $\nu$. This is quantified through the degree of second-order coherence $g^{(2)}_{\nu}(0)$, which also evolves upon propagation. The experimental results from \textbf{a} to \textbf{f} were obtained by scanning two detectors through different propagation planes. The large number of interference events upon propagation leads to the modified multiphoton system in \textbf{f}, which is characterized by a $g^{(2)}_{\nu_6} (0)$ of 1.31. This multiphoton beam exhibits quantum statistical properties that approach those observed in coherent light. The evolving quantum dynamics of our multiphoton system can be modeled through Eq. (\ref{eqn:5}). Remarkably, the conversion processes of the multiphoton system, and its modified properties of quantum coherence, take place in free space in the absence of light-matter interactions. }
   \label{Fig. 2}
\end{figure*} 

After the polarization grating, the resulting state is obtained via the transformation
\begin{equation}
    \begin{aligned}
        \hat{a}_B(\boldsymbol{\rho}) \rightarrow P_{HB}(\boldsymbol{\rho})\hat{a}_H(\boldsymbol{\rho}) + P_{VB}(\boldsymbol{\rho})\hat{a}_V(\boldsymbol{\rho}) + P_{\emptyset B}(\boldsymbol{\rho})\hat{a}_\emptyset(\boldsymbol{\rho}),
    \end{aligned}
\end{equation}
where $\hat{a}_\emptyset(\boldsymbol{\rho})$ is the mode for photon loss and $P_{AB}$ are the components of the transformation
\begin{equation}
    \boldsymbol{P}(x) = 
    \begin{bmatrix}
    \cos^2\left(\frac{\pi x}{L}\right) & \cos\left(\frac{\pi x}{L}\right)\sin\left(\frac{\pi x}{L}\right) \\
    \cos\left(\frac{\pi x}{L}\right)\sin\left(\frac{\pi x}{L}\right) & \sin^2\left(\frac{\pi x}{L}\right) \\
    \sin\left(\frac{\pi x}{L}\right) & \cos\left(\frac{\pi x}{L}\right)
    \end{bmatrix}
    \label{eqn:Prop} 
\end{equation}

where $A\in\{H,V,\emptyset\}$ and $L$ is the length of the polarization grating. The beam described by Eq. (\ref{eq:1}) is then propagated by a distance of $z$ before being measured by two pairs of photon-number-resolving (PNR) detectors \cite{you2020identification, you2021scalable}. This propagation can be modeled through the Fresnel approximation on the mode structure of the initial beam \cite{goodman2008introduction}. We can then compute the second-order correlation function \\$G^{(2)}_{ijkl} (\boldsymbol{r}_1,\boldsymbol{r}_2,z)$ for the post-selected polarization measurements in the detection plane as \cite{mandel1995optical}

\begin{equation}
    \begin{aligned}
        &G^{(2)}_{ijkl}(\boldsymbol{r}_1,\boldsymbol{r}_2,z) = \langle\hat{a}^\dagger_i(\boldsymbol{r}_1)\hat{a}^\dagger_j(\boldsymbol{r}_2)\hat{a}_k(\boldsymbol{r}_1)\hat{a}_l(\boldsymbol{r}_2)\rangle\\
        &= I_0 \int d\boldsymbol{\rho}_1 d\boldsymbol{\rho}_2 d\boldsymbol{\rho}_3 d\boldsymbol{\rho}_4 F(\boldsymbol{r}_1,\boldsymbol{r}_2,\boldsymbol{\rho}_1,\boldsymbol{\rho}_2,\boldsymbol{\rho}_3,\boldsymbol{\rho}_4,z) \\
        &\times\Big[\delta(\boldsymbol{\rho}_1-\boldsymbol{\rho}_3)\delta(\boldsymbol{\rho}_2-\boldsymbol{\rho}_4) + \delta(\boldsymbol{\rho}_1-\boldsymbol{\rho}_4)\delta(\boldsymbol{\rho}_2-\boldsymbol{\rho}_3)\Big].
    \end{aligned}
    \label{eqn:4}
\end{equation}
Remarkably, the Dirac-delta functions in Eq. (\ref{eqn:4}) demonstrate the presence of nontrivial correlations. Given the complexity of $I_0$ and $F(\boldsymbol{r}_1,\boldsymbol{r}_2,\boldsymbol{\rho}_1,\boldsymbol{\rho}_2,\boldsymbol{\rho}_3,\boldsymbol{\rho}_4,z)$, their explicit expressions are given in the SI. These describe the coherence of a photon with itself, which existed prior to interacting with the grating, and the spatial coherence gained by multiphoton scattering. These terms unveil the possibility of modifying quantum coherence of multiphoton systems upon propagation \cite{you2023multiphoton}. We use this approach to describe the correlation properties of the multiphoton wavepackets that constitute our light beam. This allows us to use an equivalent density matrix for the system $\hat{\rho}_{ijkl}(z)$ (see SI) at the detection plane to compute its corresponding joint photon-number distribution $p_{ijkl}(n_1,n_2,z)$ as
\begin{equation}
    \begin{aligned}
 p(n_1,n_2,z) = \text{Tr}\left[\hat{\rho}_{ijkl}(z) |n_1,n_2\rangle\langle n_1,n_2|\right].
    \end{aligned}
       \label{eqn:5}
\end{equation}

\begin{figure}[!t]
    \centering
\includegraphics[width=0.45\textwidth]{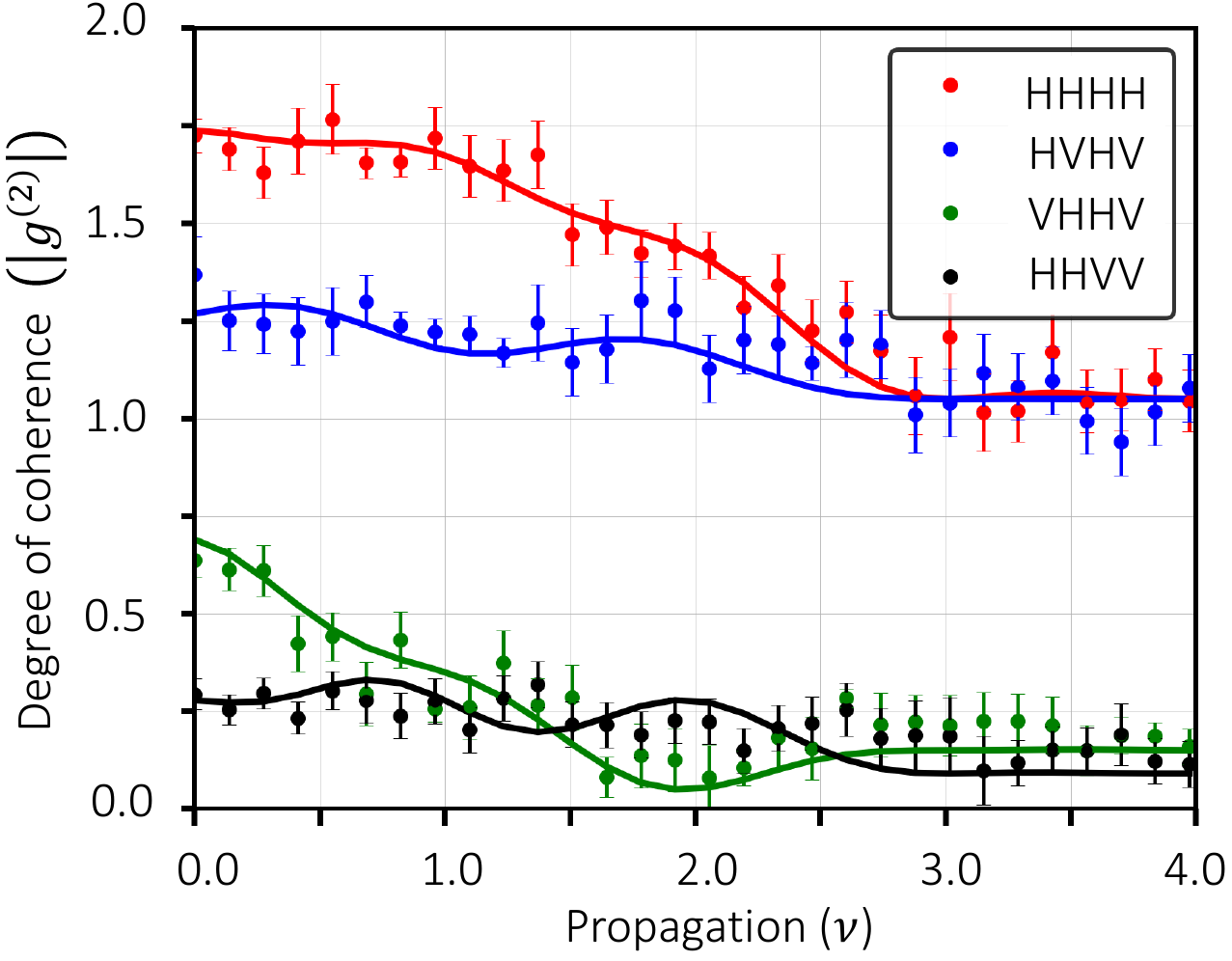}
    \caption{\textbf{Measurement of multiphoton light with sub-shot-noise properties.} We isolate multiphoton subsystems with different polarization properties. These are characterized by the degree of second-order coherence $g^{(2)}_{ijkl}$. While the four multiphoton subsystems indicate the modification of quantum coherence with the $\nu$ parameter, it should be highlighted that the subsystems described by $g^{(2)}_{\text{VHHV}}$ and $g^{(2)}_{\text{HHVV}}$ show degrees of coherence below one. Notably,  quantum light sources with quantum statistical fluctuations below the shot-noise limit show degrees of coherence smaller than one. The continuous lines represent our theoretical predictions from Eq. (\ref{eqn:4}), whereas the dots indicate experimental data.}
    \label{Fig. 3}
\end{figure}

As we shall see in the next section, these formulae allow for the prediction of very interesting correlation effects. Specifically, they predict that the statistical make-up of the light field is changing upon propagation in free space. The classical analogue to this behavior is explained by the van Cittert-Zernike theorem \cite{Cittert:34,ZERNIKE:38}, which predicts that the classical coherence properties of a light source can change upon free-space propagation. Therefore, we interpret our results in Eq. (\ref{eqn:4}) as those of a quantum van Cittert-Zernike theorem. This is because they predict the modification of quantum coherence upon free-space propagation, and that is directly analogous to the classical theorem's predictions. Specifically, Eq. (\ref{eqn:4}) predicts this free-space quantum modification through the nontrivial scattering effects induced by the Dirac-delta functions. Interestingly, these delta functions arise from the unique coherence properties of thermal light (see SI for further details). Furthermore, Eq. (\ref{eqn:5}) allows us to study multiparticle quantum coherence, which is also changing upon free-space propagation. These quantum van Cittert-Zernike effects, therefore, are not only arising in polarization subsystems, but also in multiphoton subsystems. This showcases the fundamental and intrinsic quantum impacts of free-space propagation on our state.

We explore the modification of the quantum coherence properties of propagating multiphoton systems using the experimental setup in Figure \ref{Fig. 1}\textbf{b}. We use a combination of waveplates and a spatial light modulator (SLM) to rotate the polarization properties of our multiphoton system at any transverse position \cite{PRLMirhossein2014}. In addition, this arrangement enables us to characterize the polarization and photon-number distribution of multiphoton systems at different propagation planes. Specifically, we perform measurements at different propagation planes associated with the propagation parameter $\nu=L\Delta X/(\lambda z)$. Here, the transverse distance between detectors is described by $\Delta X$ and the wavelength of the beam by $\lambda$. As demonstrated in Figure \ref{Fig. 2}, the many interference effects hosted by the propagating multiphoton system modify the photon-number distribution of the polarized components of the initial beam \cite{magana2019multiphoton, you2021observation, bhusal2021smart}. These processes lead to multiphoton systems with different quantum fluctuations and quantum properties of coherence \cite{magana2019multiphoton, Hong2023}. Each multiphoton system is characterized through the degree of second-order self coherence
\begin{equation}
    g^{(2)}_{\nu}(0)=\frac{G^{(2)}_{\text{HHHH}} (\boldsymbol{r},\boldsymbol{r},z)}{G^{(1)}_{\text{HH}}(\boldsymbol{r},z)^2},
\end{equation}
where $G^{(1)}_{i,j}(\boldsymbol{r},z) = \braket{\hat{a}^\dagger_i(\boldsymbol{r})\hat{a}_j(\boldsymbol{r})} = \sqrt{I_0}L/(2 z^2 \lambda^2)$. Interestingly, the multiphoton system in Figure \ref{Fig. 2}\textbf{a} is nearly thermal \cite{PhysRevCoherentIncoherent}. However, propagation leads to different kinds of multiphoton wavepackets. We show these from Figure \ref{Fig. 2}\textbf{a} to \textbf{f}. The coherence properties of the multiphoton system in Figure \ref{Fig. 2}\textbf{f} approach those observed in coherent light beams \cite{gerry2005introductory}. Remarkably, the conversion processes of the multiphoton system, described by Eq. (\ref{eqn:5}), take place in free space in the absence of light-matter interactions \cite{allen1987optical, venkataraman_phase_2013, Zasedatelev-2021, Snijders-2016, OLSEN2002373, you2021observation, kondakci2015}.

\begin{figure*}[!hbpt]
    \centering \includegraphics[width=\textwidth] {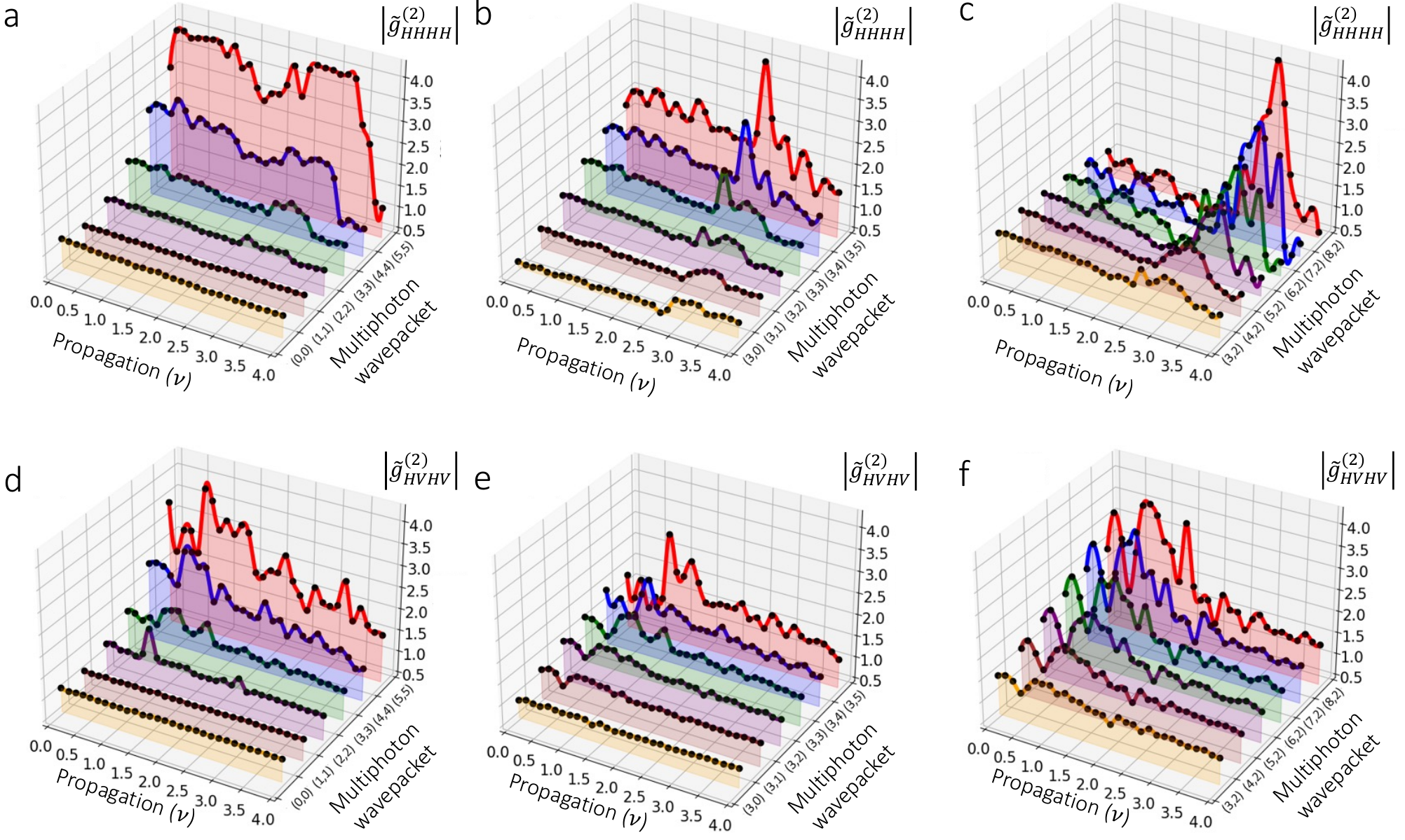}
    \caption{\textbf{Quantum coherence of propagating multiphoton wavepackets.} The panels from \textbf{a} to \textbf{c} show the evolution of multiphoton wavepackets contained in the horizontally-polarized component of the initial thermal beam. We label the multiphoton wavepacket that leads to the detection of $n_1$ photons in arm 1 and $n_2$ photons in arm 2 with $(n_1,n_2)$. The results from \textbf{a} to \textbf{c} indicate that the multiphoton events that produce the degree of second-order coherence $g_{\text{HHHH}}^{(2)}$ in Figure \ref{Fig. 3} follow distinct propagation dynamics. Although the contributions from the constituent wavepackets produce the trace described by $g_{\text{HHHH}}^{(2)}$, their individual propagation shows different coherence evolution. Specifically, we identify three representative dynamics. For example, multiphoton wavepackets with equal numbers for $n_1$ and $n_2$ exhibit the propagation dynamics in \textbf{a}. In contrast, propagating wavepackets with different values of $n_1$ and $n_2$ show a different trend for the modification of quantum coherence, these are shown in \textbf{b} and \textbf{c}. Moreover, the multiphoton wavepackets in the projected beam characterized by $g_{\text{HVHV}}^{(2)}$ exhibit the multiphoton dynamics reported from \textbf{d} to \textbf{f}. The multiphoton dynamics in these panels also depend on the number of photons in each of the measured wavepackets. }
   \label{Fig. 4}
\end{figure*} 

The polarization and photon-number properties of the propagating light beam at different transverse and longitudinal positions host many forms of interference effects \cite{ou2007multi, anno2006}. We explore these dynamics by isolating the constituent multiphoton subsystems of the propagating beam. Each multiphoton subsystem, characterized by different polarization properties, exhibits different degrees of second-order coherence \cite{bhusal2021smart}. In the experiment, we perform projective measurements on polarization. These measurements unveil the possibility of extracting multiphoton subsystems with attenuated quantum fluctuations below the shot-noise limit \cite{you2021scalable, HashemiRafsanjani2017}. In this case, we use the four detectors depicted in the experimental setup in Figure \ref{Fig. 1}\textbf{b} to perform full characterization of polarization \cite{Altepeter2005PhotonicST}. These measurements enable us to characterize correlations of multiphoton subsystems with different polarization properties, which are reported in Figure \ref{Fig. 3}. We plot the degree of second-order mutual coherence
\begin{equation}
    g^{(2)}_{ijkl}(\boldsymbol{r}_1,\boldsymbol{r}_2,z)=\frac{G^{(2)}_{ijkl}(\boldsymbol{r}_1,\boldsymbol{r}_2,z)}{G^{(1)}_{i,j}(\boldsymbol{r}_1,z)G^{(1)}_{k,l}(\boldsymbol{r}_2,z)}.
\end{equation}
The propagation of the multiphoton subsystem described by $g^{(2)}_{\text{HHHH}}$ shows a modification of the quantum statistics from super-Poissonian to nearly Poissonian \cite{Hong2023, gerry2005introductory}. A similar situation prevails for the multiphoton subsystem described by $g^{(2)}_{\text{HVHV}}$. It is worth noticing that the multiphoton subsystems described by $g^{(2)}_{\text{VHHV}}$ and $g^{(2)}_{\text{HHVV}}$ unveil the possibility of extracting multiphoton subsystems with sub-shot-noise properties \cite{agarwal2012quantum}. This implies photon-number distributions narrower than the characteristic Poissonian distribution of coherent light \cite{PhysRevCoherentIncoherent, Mandel:79}. This peculiar feature might unlock novel paths towards the implementation of sensitive measurements with sub-shot-noise fluctuations \cite{Lawrie2019SqueezedLight}. 

We now turn our attention to describe the quantum coherence evolution of propagating multiphoton wavepackets. This is explored by projecting the polarized components of the initial thermal beam into its constituent multiphoton wavepackets \cite{PhysRevCoherentIncoherent}. In this case, we analyze wavepackets with $n_1+n_2$ number of photons.  The number of photons detected in arm 1 of our experiment is described by $n_1$, whereas $n_2$ indicates the number of photons detected in arm 2. Our findings unveil that despite the fact that the degree of second-order coherence $g^{(2)}_{\text{HHHH}}$ in Figure \ref{Fig. 3} is produced by its constituent wavepackets, these show a completely different evolution of their properties of coherence. Our experimental measurements of these wavepackets can be found from Figure \ref{Fig. 4}\textbf{a} to \textbf{c}. The results in Figure \ref{Fig. 4}\textbf{a} indicate that multiphoton wavepackets, in which $n_1$ and $n_2$ are the same, show a particular evolution. In contrast, propagating wavepackets with asymmetric values of $n_1$ and $n_2$ show different trends in the modification of quantum coherence, these are shown in Figure \ref{Fig. 4}\textbf{b} and \textbf{c}. The propagation of these wavepackets can be described using Eq. (\ref{eqn:5}). Specifically, we can calculate the multiphoton degree of second-order mutual coherence \cite{dawkins2024quantum}
\begin{equation}
    \Tilde{g}^{(2)}_{ijkl}(n_1,n_2,z)=\frac{p_{ijkl}(n_1,n_2,z)}{\sum_{n}p_{ijkl}(n,n_2,z)\sum_{m}p_{ijkl}(n_1,m,z)}.
\end{equation} 
Furthermore, the multiphoton wavepackets in the projected beam, characterized by $g^{(2)}_{\text{HVHV}}$, exhibit the multiphoton dynamics reported from Figure \ref{Fig. 4}\textbf{d} to \textbf{f}. These results suggest that the multiphoton dynamics in Figure \ref{Fig. 4} depend on the number of photons in each of the measured wavepackets.

\hfill \break
This quantum field theoretic approach to studying the quantum van Cittert-Zernike theorem provides us with the ability to describe the propagation dynamics of the multiphoton systems that constitute classical light beams. We used this formalism to extract propagating multiphoton subsystems, with quantum statistical properties, from unpolarized thermal light fields. While nonlinear light-matter interactions offer the possibility of engineering complex quantum systems \cite{venkataraman_phase_2013, Zasedatelev-2021, Snijders-2016}, our scheme exploits linear propagation of multiphoton systems \cite{HallajiNatPhys2017, Mirhosseini2016PRA}. This feature enabled us to exploit multiphoton scattering in free space to produce wavepackets with different quantum statistical properties \cite{magana2019multiphoton}. As such, our work combines the benefits of post-selective measurements with those of multiphoton scattering in propagating light beams, and it allows us to study the modification of the quantum statistical properties of multiphoton wavepackets in free space. Although, the incoherent combination of light beams with different polarization properties can lead to the modification of the degree of second-order coherence \cite{you2021observation,Hong2024NP}, we performed direct measurements of polarized multiphoton systems with propagating quantum coherence properties (see Fig. \ref{Fig. 4}\textbf{a} to \textbf{c}). Interestingly, these processes are defined by the number of particles in the measured multiphoton system. Consequently, these findings have important implications for all-optical engineering of multiphoton quantum systems.

\section{Conclusion}
We demonstrated the possibility of modifying the excitation mode of thermal multiphoton fields through free space propagation. This modification stems from the scattering of multiphoton wavepackets in the absence of light-matter interactions \cite{allen1987optical, venkataraman_phase_2013, Zasedatelev-2021, Snijders-2016, OLSEN2002373, you2021observation, kondakci2015,Wen:13}. The modification of the excitation mode of a photonic system and its associated quantum fluctuations result in the formation of different light fields with distinct quantum coherence properties \cite{PhysRevCoherentIncoherent, GlauberPR1963, PhysRevLettSudarshan}. The evolution of multiphoton quantum coherence is described through the nonclassical formulation of the van Cittert-Zernike theorem, unveiling conditions for the formation of multiphoton systems with attenuated quantum fluctuations below the sub-shot-noise limit \cite{you2021scalable,Batarseh:18, Lawrie2019SqueezedLight}. Notably, these quantum multiphoton systems emerge in the absence of optical nonlinearities, suggesting an all-optical approach for extracting multiphoton wavepackets with nonclassical statistics. We believe that the identification of this surprising multiphoton dynamics has important implications for multiphoton protocols quantum information \cite{Walmsley:23, aspuru-guzik_photonic_2012}.

\section*{Acknowledgments}
J.F. acknowledges funding from the National Science Foundation through Grant No. OMA 2231387. M.H., R.B.D., C.Y. and O.S.M.L. acknowledge support from the Army Research Office (ARO), through the Early Career Program (ECP) under the grant no. W911NF-22-1-0088. R.J.L.-M. thankfully acknowledges financial support by DGAPA-UNAM under the project UNAM-PAPIIT IN101623. 



\section*{References}
\bibliography{main} 

\begin{thebibliography}{50}%
\makeatletter
\providecommand \@ifxundefined [1]{%
 \@ifx{#1\undefined}
}%
\providecommand \@ifnum [1]{%
 \ifnum #1\expandafter \@firstoftwo
 \else \expandafter \@secondoftwo
 \fi
}%
\providecommand \@ifx [1]{%
 \ifx #1\expandafter \@firstoftwo
 \else \expandafter \@secondoftwo
 \fi
}%
\providecommand \natexlab [1]{#1}%
\providecommand \enquote  [1]{``#1''}%
\providecommand \bibnamefont  [1]{#1}%
\providecommand \bibfnamefont [1]{#1}%
\providecommand \citenamefont [1]{#1}%
\providecommand \href@noop [0]{\@secondoftwo}%
\providecommand \href [0]{\begingroup \@sanitize@url \@href}%
\providecommand \@href[1]{\@@startlink{#1}\@@href}%
\providecommand \@@href[1]{\endgroup#1\@@endlink}%
\providecommand \@sanitize@url [0]{\catcode `\\12\catcode `\$12\catcode
  `\&12\catcode `\#12\catcode `\^12\catcode `\_12\catcode `\%12\relax}%
\providecommand \@@startlink[1]{}%
\providecommand \@@endlink[0]{}%
\providecommand \url  [0]{\begingroup\@sanitize@url \@url }%
\providecommand \@url [1]{\endgroup\@href {#1}{\urlprefix }}%
\providecommand \urlprefix  [0]{URL }%
\providecommand \Eprint [0]{\href }%
\providecommand \doibase [0]{https://doi.org/}%
\providecommand \selectlanguage [0]{\@gobble}%
\providecommand \bibinfo  [0]{\@secondoftwo}%
\providecommand \bibfield  [0]{\@secondoftwo}%
\providecommand \translation [1]{[#1]}%
\providecommand \BibitemOpen [0]{}%
\providecommand \bibitemStop [0]{}%
\providecommand \bibitemNoStop [0]{.\EOS\space}%
\providecommand \EOS [0]{\spacefactor3000\relax}%
\providecommand \BibitemShut  [1]{\csname bibitem#1\endcsname}%
\let\auto@bib@innerbib\@empty
\bibitem [{\citenamefont {Wolf}(1954)}]{wolf1954optics}%
  \BibitemOpen
  \bibfield  {author} {\bibinfo {author} {\bibfnamefont {E.}~\bibnamefont
  {Wolf}},\ }\bibfield  {title} {\bibinfo {title} {Optics in terms of
  observable quantities},\ }\href@noop {} {\bibfield  {journal} {\bibinfo
  {journal} {Il Nuovo Cimento (1943-1954)}\ }\textbf {\bibinfo {volume} {12}},\
  \bibinfo {pages} {884} (\bibinfo {year} {1954})}\BibitemShut {NoStop}%
\bibitem [{\citenamefont {Mandel}\ and\ \citenamefont
  {Wolf}(1995)}]{mandel1995optical}%
  \BibitemOpen
  \bibfield  {author} {\bibinfo {author} {\bibfnamefont {L.}~\bibnamefont
  {Mandel}}\ and\ \bibinfo {author} {\bibfnamefont {E.}~\bibnamefont {Wolf}},\
  }\href@noop {} {\emph {\bibinfo {title} {Optical coherence and quantum
  optics}}}\ (\bibinfo  {publisher} {Cambridge university press},\ \bibinfo
  {year} {1995})\BibitemShut {NoStop}%
\bibitem [{\citenamefont {Born}\ and\ \citenamefont
  {Wolf}(2013)}]{born2013principles}%
  \BibitemOpen
  \bibfield  {author} {\bibinfo {author} {\bibfnamefont {M.}~\bibnamefont
  {Born}}\ and\ \bibinfo {author} {\bibfnamefont {E.}~\bibnamefont {Wolf}},\
  }\href@noop {} {\emph {\bibinfo {title} {Principles of optics:
  electromagnetic theory of propagation, interference and diffraction of
  light}}}\ (\bibinfo  {publisher} {Elsevier},\ \bibinfo {year}
  {2013})\BibitemShut {NoStop}%
\bibitem [{\citenamefont {Dorrer}(2004)}]{Dorrer:04}%
  \BibitemOpen
  \bibfield  {author} {\bibinfo {author} {\bibfnamefont {C.}~\bibnamefont
  {Dorrer}},\ }\bibfield  {title} {\bibinfo {title} {Temporal van
  {C}ittert-{Z}ernike theorem and its application to the measurement of
  chromatic dispersion},\ }\href {https://doi.org/10.1364/JOSAB.21.001417}
  {\bibfield  {journal} {\bibinfo  {journal} {J. Opt. Soc. Am. B}\ }\textbf
  {\bibinfo {volume} {21}},\ \bibinfo {pages} {1417\textendash1423} (\bibinfo
  {year} {2004})}\BibitemShut {NoStop}%
\bibitem [{\citenamefont {Gori}\ \emph {et~al.}(2000)\citenamefont {Gori},
  \citenamefont {Santarsiero}, \citenamefont {Borghi},\ and\ \citenamefont
  {Piquero}}]{gori2000use}%
  \BibitemOpen
  \bibfield  {author} {\bibinfo {author} {\bibfnamefont {F.}~\bibnamefont
  {Gori}}, \bibinfo {author} {\bibfnamefont {M.}~\bibnamefont {Santarsiero}},
  \bibinfo {author} {\bibfnamefont {R.}~\bibnamefont {Borghi}},\ and\ \bibinfo
  {author} {\bibfnamefont {G.}~\bibnamefont {Piquero}},\ }\bibfield  {title}
  {\bibinfo {title} {Use of the van {C}ittert--{Z}ernike theorem for partially
  polarized sources},\ }\href {https://doi.org/10.1364/OL.25.001291} {\bibfield
   {journal} {\bibinfo  {journal} {Opt. Lett.}\ }\textbf {\bibinfo {volume}
  {25}},\ \bibinfo {pages} {1291\textendash1293} (\bibinfo {year}
  {2000})}\BibitemShut {NoStop}%
\bibitem [{\citenamefont {Zhang}\ \emph {et~al.}(2020)\citenamefont {Zhang},
  \citenamefont {Cai},\ and\ \citenamefont {Gbur}}]{Cai2020}%
  \BibitemOpen
  \bibfield  {author} {\bibinfo {author} {\bibfnamefont {Y.}~\bibnamefont
  {Zhang}}, \bibinfo {author} {\bibfnamefont {Y.}~\bibnamefont {Cai}},\ and\
  \bibinfo {author} {\bibfnamefont {G.}~\bibnamefont {Gbur}},\ }\bibfield
  {title} {\bibinfo {title} {Partially coherent vortex beams of arbitrary
  radial order and a van {C}ittert--{Z}ernike theorem for vortices},\ }\href
  {https://doi.org/10.1103/PhysRevA.101.043812} {\bibfield  {journal} {\bibinfo
   {journal} {Phys. Rev. A}\ }\textbf {\bibinfo {volume} {101}},\ \bibinfo
  {pages} {043812} (\bibinfo {year} {2020})}\BibitemShut {NoStop}%
\bibitem [{\citenamefont {Cai}\ \emph {et~al.}(2012)\citenamefont {Cai},
  \citenamefont {Dong},\ and\ \citenamefont {Hoenders}}]{Cai2012}%
  \BibitemOpen
  \bibfield  {author} {\bibinfo {author} {\bibfnamefont {Y.}~\bibnamefont
  {Cai}}, \bibinfo {author} {\bibfnamefont {Y.}~\bibnamefont {Dong}},\ and\
  \bibinfo {author} {\bibfnamefont {B.}~\bibnamefont {Hoenders}},\ }\bibfield
  {title} {\bibinfo {title} {Interdependence between the temporal and spatial
  longitudinal and transverse degrees of partial coherence and a generalization
  of the van {C}ittert-{Z}ernike theorem},\ }\href
  {https://doi.org/10.1364/JOSAA.29.002542} {\bibfield  {journal} {\bibinfo
  {journal} {J. Opt. Soc. Am. A}\ }\textbf {\bibinfo {volume} {29}},\ \bibinfo
  {pages} {2542\textendash2551} (\bibinfo {year} {2012})}\BibitemShut {NoStop}%
\bibitem [{\citenamefont {Saleh}\ \emph {et~al.}(2005)\citenamefont {Saleh},
  \citenamefont {Teich},\ and\ \citenamefont {Sergienko}}]{Saleh05PRL}%
  \BibitemOpen
  \bibfield  {author} {\bibinfo {author} {\bibfnamefont {B.~E.~A.}\
  \bibnamefont {Saleh}}, \bibinfo {author} {\bibfnamefont {M.~C.}\ \bibnamefont
  {Teich}},\ and\ \bibinfo {author} {\bibfnamefont {A.~V.}\ \bibnamefont
  {Sergienko}},\ }\bibfield  {title} {\bibinfo {title} {Wolf equations for
  two-photon light},\ }\href {https://doi.org/10.1103/PhysRevLett.94.223601}
  {\bibfield  {journal} {\bibinfo  {journal} {Phys. Rev. Lett.}\ }\textbf
  {\bibinfo {volume} {94}},\ \bibinfo {pages} {223601} (\bibinfo {year}
  {2005})}\BibitemShut {NoStop}%
\bibitem [{\citenamefont {You}\ \emph {et~al.}(2023)\citenamefont {You},
  \citenamefont {Miller}, \citenamefont {Le{\'o}n-Montiel},\ and\ \citenamefont
  {Maga{\~n}a-Loaiza}}]{you2023multiphoton}%
  \BibitemOpen
  \bibfield  {author} {\bibinfo {author} {\bibfnamefont {C.}~\bibnamefont
  {You}}, \bibinfo {author} {\bibfnamefont {A.}~\bibnamefont {Miller}},
  \bibinfo {author} {\bibfnamefont {R.~d.~J.}\ \bibnamefont
  {Le{\'o}n-Montiel}},\ and\ \bibinfo {author} {\bibfnamefont {O.~S.}\
  \bibnamefont {Maga{\~n}a-Loaiza}},\ }\bibfield  {title} {\bibinfo {title}
  {Multiphoton quantum van cittert-zernike theorem},\ }\href
  {https://doi.org/10.1038/s41534-023-00720-w} {\bibfield  {journal} {\bibinfo
  {journal} {npj Quantum Inf.}\ }\textbf {\bibinfo {volume} {9}},\ \bibinfo
  {pages} {50} (\bibinfo {year} {2023})}\BibitemShut {NoStop}%
\bibitem [{\citenamefont {{van Cittert}}(1934)}]{Cittert:34}%
  \BibitemOpen
  \bibfield  {author} {\bibinfo {author} {\bibfnamefont {P.}~\bibnamefont {{van
  Cittert}}},\ }\bibfield  {title} {\bibinfo {title} {Die wahrscheinliche
  schwingungsverteilung in einer von einer lichtquelle direkt oder mittels
  einer linse beleuchteten ebene},\ }\href
  {https://doi.org/https://doi.org/10.1016/S0031-8914(34)90026-4} {\bibfield
  {journal} {\bibinfo  {journal} {Physica}\ }\textbf {\bibinfo {volume} {1}},\
  \bibinfo {pages} {201\textendash210} (\bibinfo {year} {1934})}\BibitemShut
  {NoStop}%
\bibitem [{\citenamefont {Zernike}(1938)}]{ZERNIKE:38}%
  \BibitemOpen
  \bibfield  {author} {\bibinfo {author} {\bibfnamefont {F.}~\bibnamefont
  {Zernike}},\ }\bibfield  {title} {\bibinfo {title} {The concept of degree of
  coherence and its application to optical problems},\ }\href
  {https://doi.org/https://doi.org/10.1016/S0031-8914(38)80203-2} {\bibfield
  {journal} {\bibinfo  {journal} {Physica}\ }\textbf {\bibinfo {volume} {5}},\
  \bibinfo {pages} {785\textendash795} (\bibinfo {year} {1938})}\BibitemShut
  {NoStop}%
\bibitem [{\citenamefont {Glauber}(1963{\natexlab{a}})}]{GlauberPR1963}%
  \BibitemOpen
  \bibfield  {author} {\bibinfo {author} {\bibfnamefont {R.~J.}\ \bibnamefont
  {Glauber}},\ }\bibfield  {title} {\bibinfo {title} {The quantum theory of
  optical coherence},\ }\href {https://doi.org/10.1103/PhysRev.130.2529}
  {\bibfield  {journal} {\bibinfo  {journal} {Phys. Rev.}\ }\textbf {\bibinfo
  {volume} {130}},\ \bibinfo {pages} {2529} (\bibinfo {year}
  {1963}{\natexlab{a}})}\BibitemShut {NoStop}%
\bibitem [{\citenamefont {Sudarshan}(1963)}]{PhysRevLettSudarshan}%
  \BibitemOpen
  \bibfield  {author} {\bibinfo {author} {\bibfnamefont {E.~C.~G.}\
  \bibnamefont {Sudarshan}},\ }\bibfield  {title} {\bibinfo {title}
  {Equivalence of semiclassical and quantum mechanical descriptions of
  statistical light beams},\ }\href
  {https://doi.org/10.1103/PhysRevLett.10.277} {\bibfield  {journal} {\bibinfo
  {journal} {Phys. Rev. Lett.}\ }\textbf {\bibinfo {volume} {10}},\ \bibinfo
  {pages} {277} (\bibinfo {year} {1963})}\BibitemShut {NoStop}%
\bibitem [{\citenamefont {Dell'Anno}\ \emph {et~al.}(2006)\citenamefont
  {Dell'Anno}, \citenamefont {Siena},\ and\ \citenamefont
  {Illuminati}}]{anno2006}%
  \BibitemOpen
  \bibfield  {author} {\bibinfo {author} {\bibfnamefont {F.}~\bibnamefont
  {Dell'Anno}}, \bibinfo {author} {\bibfnamefont {S.~D.}\ \bibnamefont
  {Siena}},\ and\ \bibinfo {author} {\bibfnamefont {F.}~\bibnamefont
  {Illuminati}},\ }\bibfield  {title} {\bibinfo {title} {Multiphoton quantum
  optics and quantum state engineering},\ }\href
  {https://doi.org/10.1016/j.physrep.2006.01.004} {\bibfield  {journal}
  {\bibinfo  {journal} {Phys. Rep.}\ }\textbf {\bibinfo {volume} {428}},\
  \bibinfo {pages} {53} (\bibinfo {year} {2006})}\BibitemShut {NoStop}%
\bibitem [{\citenamefont {Maga\~na Loaiza}\ and\ \citenamefont
  {et~al.}(2019)}]{magana2019multiphoton}%
  \BibitemOpen
  \bibfield  {author} {\bibinfo {author} {\bibfnamefont {O.~S.}\ \bibnamefont
  {Maga\~na Loaiza}}\ and\ \bibinfo {author} {\bibnamefont {et~al.}},\
  }\bibfield  {title} {\bibinfo {title} {Multiphoton quantum-state engineering
  using conditional measurements},\ }\href
  {https://doi.org/10.1038/s41534-019-0195-2} {\bibfield  {journal} {\bibinfo
  {journal} {npj Quantum Inf.}\ }\textbf {\bibinfo {volume} {5}},\ \bibinfo
  {pages} {80} (\bibinfo {year} {2019})}\BibitemShut {NoStop}%
\bibitem [{\citenamefont {You}\ \emph {et~al.}(2020)\citenamefont {You},
  \citenamefont {Nellikka}, \citenamefont {Leon},\ and\ \citenamefont
  {Magaña-Loaiza}}]{You2020plasmonics}%
  \BibitemOpen
  \bibfield  {author} {\bibinfo {author} {\bibfnamefont {C.}~\bibnamefont
  {You}}, \bibinfo {author} {\bibfnamefont {A.~C.}\ \bibnamefont {Nellikka}},
  \bibinfo {author} {\bibfnamefont {I.~D.}\ \bibnamefont {Leon}},\ and\
  \bibinfo {author} {\bibfnamefont {O.~S.}\ \bibnamefont {Magaña-Loaiza}},\
  }\bibfield  {title} {\bibinfo {title} {Multiparticle quantum plasmonics},\
  }\href {https://doi.org/doi:10.1515/nanoph-2019-0517} {\bibfield  {journal}
  {\bibinfo  {journal} {Nanophotonics}\ }\textbf {\bibinfo {volume} {9}},\
  \bibinfo {pages} {1243\textendash1269} (\bibinfo {year} {2020})}\BibitemShut
  {NoStop}%
\bibitem [{\citenamefont {You}\ and\ \citenamefont
  {et~al.}(2021{\natexlab{a}})}]{you2021observation}%
  \BibitemOpen
  \bibfield  {author} {\bibinfo {author} {\bibfnamefont {C.}~\bibnamefont
  {You}}\ and\ \bibinfo {author} {\bibnamefont {et~al.}},\ }\bibfield  {title}
  {\bibinfo {title} {Observation of the modification of quantum statistics of
  plasmonic systems},\ }\href
  {https://doi.org/https://doi.org/10.1038/s41467-021-25489-4} {\bibfield
  {journal} {\bibinfo  {journal} {Nat. Commun.}\ }\textbf {\bibinfo {volume}
  {12}},\ \bibinfo {pages} {5161} (\bibinfo {year}
  {2021}{\natexlab{a}})}\BibitemShut {NoStop}%
\bibitem [{\citenamefont {Walmsley}(2023)}]{Walmsley:23}%
  \BibitemOpen
  \bibfield  {author} {\bibinfo {author} {\bibfnamefont {I.}~\bibnamefont
  {Walmsley}},\ }\bibfield  {title} {\bibinfo {title} {Light in quantum
  computing and simulation: perspective},\ }\href
  {https://doi.org/10.1364/OPTICAQ.507527} {\bibfield  {journal} {\bibinfo
  {journal} {Optica Quantum}\ }\textbf {\bibinfo {volume} {1}},\ \bibinfo
  {pages} {35} (\bibinfo {year} {2023})}\BibitemShut {NoStop}%
\bibitem [{\citenamefont {Maga{\~{n}}a-Loaiza}\ and\ \citenamefont
  {Boyd}(2019)}]{Omar2019Review}%
  \BibitemOpen
  \bibfield  {author} {\bibinfo {author} {\bibfnamefont {O.~S.}\ \bibnamefont
  {Maga{\~{n}}a-Loaiza}}\ and\ \bibinfo {author} {\bibfnamefont {R.~W.}\
  \bibnamefont {Boyd}},\ }\bibfield  {title} {\bibinfo {title} {Quantum imaging
  and information},\ }\href {https://doi.org/10.1088/1361-6633/ab5005}
  {\bibfield  {journal} {\bibinfo  {journal} {Rep. Prog. Phys.}\ }\textbf
  {\bibinfo {volume} {82}},\ \bibinfo {pages} {124401} (\bibinfo {year}
  {2019})}\BibitemShut {NoStop}%
\bibitem [{\citenamefont {Bhusal}\ and\ \citenamefont
  {et~al.}(2022)}]{bhusal2021smart}%
  \BibitemOpen
  \bibfield  {author} {\bibinfo {author} {\bibfnamefont {N.}~\bibnamefont
  {Bhusal}}\ and\ \bibinfo {author} {\bibnamefont {et~al.}},\ }\bibfield
  {title} {\bibinfo {title} {Smart quantum statistical imaging beyond the
  {Abbe}-{Rayleigh} criterion},\ }\href
  {https://doi.org/10.1038/s41534-022-00593-5} {\bibfield  {journal} {\bibinfo
  {journal} {npj Quantum Inf.}\ }\textbf {\bibinfo {volume} {8}},\ \bibinfo
  {pages} {83} (\bibinfo {year} {2022})}\BibitemShut {NoStop}%
\bibitem [{\citenamefont {Aspuru-Guzik}\ and\ \citenamefont
  {Walther}(2012)}]{aspuru-guzik_photonic_2012}%
  \BibitemOpen
  \bibfield  {author} {\bibinfo {author} {\bibfnamefont {A.}~\bibnamefont
  {Aspuru-Guzik}}\ and\ \bibinfo {author} {\bibfnamefont {P.}~\bibnamefont
  {Walther}},\ }\bibfield  {title} {\bibinfo {title} {Photonic quantum
  simulators},\ }\href {https://doi.org/10.1038/nphys2253} {\bibfield
  {journal} {\bibinfo  {journal} {Nat. Phys.}\ }\textbf {\bibinfo {volume}
  {8}},\ \bibinfo {pages} {285\textendash291} (\bibinfo {year}
  {2012})}\BibitemShut {NoStop}%
\bibitem [{\citenamefont
  {Glauber}(1963{\natexlab{b}})}]{PhysRevCoherentIncoherent}%
  \BibitemOpen
  \bibfield  {author} {\bibinfo {author} {\bibfnamefont {R.~J.}\ \bibnamefont
  {Glauber}},\ }\bibfield  {title} {\bibinfo {title} {Coherent and incoherent
  states of the radiation field},\ }\href
  {https://doi.org/10.1103/PhysRev.131.2766} {\bibfield  {journal} {\bibinfo
  {journal} {Phys. Rev.}\ }\textbf {\bibinfo {volume} {131}},\ \bibinfo {pages}
  {2766} (\bibinfo {year} {1963}{\natexlab{b}})}\BibitemShut {NoStop}%
\bibitem [{\citenamefont {You}\ and\ \citenamefont
  {et~al.}(2020)}]{you2020identification}%
  \BibitemOpen
  \bibfield  {author} {\bibinfo {author} {\bibfnamefont {C.}~\bibnamefont
  {You}}\ and\ \bibinfo {author} {\bibnamefont {et~al.}},\ }\bibfield  {title}
  {\bibinfo {title} {Identification of light sources using machine learning},\
  }\href {https://doi.org/10.1063/1.5133846} {\bibfield  {journal} {\bibinfo
  {journal} {Appl. Phys. Rev.}\ }\textbf {\bibinfo {volume} {7}},\ \bibinfo
  {pages} {021404} (\bibinfo {year} {2020})}\BibitemShut {NoStop}%
\bibitem [{\citenamefont {Kong}\ \emph {et~al.}(2024)\citenamefont {Kong},
  \citenamefont {Zhang}, \citenamefont {Liu}, \citenamefont {Li}, \citenamefont
  {Wang}, \citenamefont {Xie},\ and\ \citenamefont {You}}]{YouLXAP2024}%
  \BibitemOpen
  \bibfield  {author} {\bibinfo {author} {\bibfnamefont {L.-D.}\ \bibnamefont
  {Kong}}, \bibinfo {author} {\bibfnamefont {T.-Z.}\ \bibnamefont {Zhang}},
  \bibinfo {author} {\bibfnamefont {X.-Y.}\ \bibnamefont {Liu}}, \bibinfo
  {author} {\bibfnamefont {H.}~\bibnamefont {Li}}, \bibinfo {author}
  {\bibfnamefont {Z.}~\bibnamefont {Wang}}, \bibinfo {author} {\bibfnamefont
  {X.-M.}\ \bibnamefont {Xie}},\ and\ \bibinfo {author} {\bibfnamefont {L.-X.}\
  \bibnamefont {You}},\ }\bibfield  {title} {\bibinfo {title}
  {{Large-inductance superconducting microstrip photon detector enabling 10
  photon-number resolution}},\ }\href {https://doi.org/10.1117/1.AP.6.1.016004}
  {\bibfield  {journal} {\bibinfo  {journal} {Advanced Photonics}\ }\textbf
  {\bibinfo {volume} {6}},\ \bibinfo {pages} {016004} (\bibinfo {year}
  {2024})}\BibitemShut {NoStop}%
\bibitem [{\citenamefont {Allen}\ and\ \citenamefont
  {Eberly}(1987)}]{allen1987optical}%
  \BibitemOpen
  \bibfield  {author} {\bibinfo {author} {\bibfnamefont {L.}~\bibnamefont
  {Allen}}\ and\ \bibinfo {author} {\bibfnamefont {J.~H.}\ \bibnamefont
  {Eberly}},\ }\href@noop {} {\emph {\bibinfo {title} {Optical resonance and
  two-level atoms}}},\ Vol.~\bibinfo {volume} {28}\ (\bibinfo  {publisher}
  {Courier Corporation},\ \bibinfo {year} {1987})\BibitemShut {NoStop}%
\bibitem [{\citenamefont {Venkataraman}\ \emph {et~al.}(2013)\citenamefont
  {Venkataraman}, \citenamefont {Saha},\ and\ \citenamefont
  {Gaeta}}]{venkataraman_phase_2013}%
  \BibitemOpen
  \bibfield  {author} {\bibinfo {author} {\bibfnamefont {V.}~\bibnamefont
  {Venkataraman}}, \bibinfo {author} {\bibfnamefont {K.}~\bibnamefont {Saha}},\
  and\ \bibinfo {author} {\bibfnamefont {A.~L.}\ \bibnamefont {Gaeta}},\
  }\bibfield  {title} {\bibinfo {title} {Phase modulation at the few-photon
  level for weak-nonlinearity-based quantum computing},\ }\href
  {https://doi.org/10.1038/nphoton.2012.283} {\bibfield  {journal} {\bibinfo
  {journal} {Nat. Photonics}\ }\textbf {\bibinfo {volume} {7}},\ \bibinfo
  {pages} {138\textendash141} (\bibinfo {year} {2013})}\BibitemShut {NoStop}%
\bibitem [{\citenamefont {Zasedatelev}\ \emph {et~al.}(2021)\citenamefont
  {Zasedatelev}, \citenamefont {Baranikov}, \citenamefont {Sannikov},
  \citenamefont {Urbonas}, \citenamefont {Scafirimuto}, \citenamefont
  {Shishkov}, \citenamefont {Andrianov}, \citenamefont {Lozovik}, \citenamefont
  {Scherf}, \citenamefont {St{\"o}ferle}, \citenamefont {Mahrt},\ and\
  \citenamefont {Lagoudakis}}]{Zasedatelev-2021}%
  \BibitemOpen
  \bibfield  {author} {\bibinfo {author} {\bibfnamefont {A.~V.}\ \bibnamefont
  {Zasedatelev}}, \bibinfo {author} {\bibfnamefont {A.~V.}\ \bibnamefont
  {Baranikov}}, \bibinfo {author} {\bibfnamefont {D.}~\bibnamefont {Sannikov}},
  \bibinfo {author} {\bibfnamefont {D.}~\bibnamefont {Urbonas}}, \bibinfo
  {author} {\bibfnamefont {F.}~\bibnamefont {Scafirimuto}}, \bibinfo {author}
  {\bibfnamefont {V.~Y.}\ \bibnamefont {Shishkov}}, \bibinfo {author}
  {\bibfnamefont {E.~S.}\ \bibnamefont {Andrianov}}, \bibinfo {author}
  {\bibfnamefont {Y.~E.}\ \bibnamefont {Lozovik}}, \bibinfo {author}
  {\bibfnamefont {U.}~\bibnamefont {Scherf}}, \bibinfo {author} {\bibfnamefont
  {T.}~\bibnamefont {St{\"o}ferle}}, \bibinfo {author} {\bibfnamefont {R.~F.}\
  \bibnamefont {Mahrt}},\ and\ \bibinfo {author} {\bibfnamefont {P.~G.}\
  \bibnamefont {Lagoudakis}},\ }\bibfield  {title} {\bibinfo {title}
  {Single-photon nonlinearity at room temperature},\ }\href
  {https://doi.org/10.1038/s41586-021-03866-9} {\bibfield  {journal} {\bibinfo
  {journal} {Nature}\ }\textbf {\bibinfo {volume} {597}},\ \bibinfo {pages}
  {493} (\bibinfo {year} {2021})}\BibitemShut {NoStop}%
\bibitem [{\citenamefont {Snijders}\ \emph {et~al.}(2016)\citenamefont
  {Snijders}, \citenamefont {Frey}, \citenamefont {Norman}, \citenamefont
  {Bakker}, \citenamefont {Langman}, \citenamefont {Gossard}, \citenamefont
  {Bowers}, \citenamefont {van Exter}, \citenamefont {Bouwmeester},\ and\
  \citenamefont {L{\"o}ffler}}]{Snijders-2016}%
  \BibitemOpen
  \bibfield  {author} {\bibinfo {author} {\bibfnamefont {H.}~\bibnamefont
  {Snijders}}, \bibinfo {author} {\bibfnamefont {J.~A.}\ \bibnamefont {Frey}},
  \bibinfo {author} {\bibfnamefont {J.}~\bibnamefont {Norman}}, \bibinfo
  {author} {\bibfnamefont {M.~P.}\ \bibnamefont {Bakker}}, \bibinfo {author}
  {\bibfnamefont {E.~C.}\ \bibnamefont {Langman}}, \bibinfo {author}
  {\bibfnamefont {A.}~\bibnamefont {Gossard}}, \bibinfo {author} {\bibfnamefont
  {J.~E.}\ \bibnamefont {Bowers}}, \bibinfo {author} {\bibfnamefont {M.~P.}\
  \bibnamefont {van Exter}}, \bibinfo {author} {\bibfnamefont {D.}~\bibnamefont
  {Bouwmeester}},\ and\ \bibinfo {author} {\bibfnamefont {W.}~\bibnamefont
  {L{\"o}ffler}},\ }\bibfield  {title} {\bibinfo {title} {Purification of a
  single-photon nonlinearity},\ }\href {https://doi.org/10.1038/ncomms12578}
  {\bibfield  {journal} {\bibinfo  {journal} {Nat. Commun.}\ }\textbf {\bibinfo
  {volume} {7}},\ \bibinfo {pages} {12578} (\bibinfo {year}
  {2016})}\BibitemShut {NoStop}%
\bibitem [{\citenamefont {Hallaji}\ \emph {et~al.}(2017)\citenamefont
  {Hallaji}, \citenamefont {Feizpour}, \citenamefont {Dmochowski},
  \citenamefont {Sinclair},\ and\ \citenamefont
  {Steinberg}}]{HallajiNatPhys2017}%
  \BibitemOpen
  \bibfield  {author} {\bibinfo {author} {\bibfnamefont {M.}~\bibnamefont
  {Hallaji}}, \bibinfo {author} {\bibfnamefont {A.}~\bibnamefont {Feizpour}},
  \bibinfo {author} {\bibfnamefont {G.}~\bibnamefont {Dmochowski}}, \bibinfo
  {author} {\bibfnamefont {J.}~\bibnamefont {Sinclair}},\ and\ \bibinfo
  {author} {\bibfnamefont {A.~M.}\ \bibnamefont {Steinberg}},\ }\bibfield
  {title} {\bibinfo {title} {Weak-value amplification of the nonlinear effect
  of a single photon},\ }\href {https://doi.org/10.1038/nphys4040} {\bibfield
  {journal} {\bibinfo  {journal} {Nature Physics}\ }\textbf {\bibinfo {volume}
  {13}},\ \bibinfo {pages} {540} (\bibinfo {year} {2017})}\BibitemShut
  {NoStop}%
\bibitem [{\citenamefont {Mirhosseini}\ \emph {et~al.}(2014)\citenamefont
  {Mirhosseini}, \citenamefont {Maga\~na Loaiza}, \citenamefont
  {Hashemi~Rafsanjani},\ and\ \citenamefont {Boyd}}]{PRLMirhossein2014}%
  \BibitemOpen
  \bibfield  {author} {\bibinfo {author} {\bibfnamefont {M.}~\bibnamefont
  {Mirhosseini}}, \bibinfo {author} {\bibfnamefont {O.~S.}\ \bibnamefont
  {Maga\~na Loaiza}}, \bibinfo {author} {\bibfnamefont {S.~M.}\ \bibnamefont
  {Hashemi~Rafsanjani}},\ and\ \bibinfo {author} {\bibfnamefont {R.~W.}\
  \bibnamefont {Boyd}},\ }\bibfield  {title} {\bibinfo {title} {Compressive
  direct measurement of the quantum wave function},\ }\href
  {https://doi.org/10.1103/PhysRevLett.113.090402} {\bibfield  {journal}
  {\bibinfo  {journal} {Phys. Rev. Lett.}\ }\textbf {\bibinfo {volume} {113}},\
  \bibinfo {pages} {090402} (\bibinfo {year} {2014})}\BibitemShut {NoStop}%
\bibitem [{\citenamefont {Altepeter}\ \emph {et~al.}(2005)\citenamefont
  {Altepeter}, \citenamefont {Jeffrey},\ and\ \citenamefont
  {Kwiat}}]{Altepeter2005PhotonicST}%
  \BibitemOpen
  \bibfield  {author} {\bibinfo {author} {\bibfnamefont {J.}~\bibnamefont
  {Altepeter}}, \bibinfo {author} {\bibfnamefont {E.}~\bibnamefont {Jeffrey}},\
  and\ \bibinfo {author} {\bibfnamefont {P.}~\bibnamefont {Kwiat}},\ }\bibfield
   {title} {\bibinfo {title} {Photonic state tomography},\ }\href
  {https://doi.org/https://doi.org/10.1016/S1049-250X(05)52003-2} {\bibfield
  {journal} {\bibinfo  {journal} {Adv. At. Mol. Opt. Phys.}\ }\textbf {\bibinfo
  {volume} {52}},\ \bibinfo {pages} {105} (\bibinfo {year} {2005})}\BibitemShut
  {NoStop}%
\bibitem [{\citenamefont {Hashemi~Rafsanjani}\ \emph
  {et~al.}(2017)\citenamefont {Hashemi~Rafsanjani}, \citenamefont
  {Mirhosseini}, \citenamefont {Magana-Loaiza}, \citenamefont {Gard},
  \citenamefont {Birrittella}, \citenamefont {Koltenbah}, \citenamefont
  {Parazzoli}, \citenamefont {Capron}, \citenamefont {Gerry}, \citenamefont
  {Dowling},\ and\ \citenamefont {Boyd}}]{HashemiRafsanjani2017}%
  \BibitemOpen
  \bibfield  {author} {\bibinfo {author} {\bibfnamefont {S.~M.}\ \bibnamefont
  {Hashemi~Rafsanjani}}, \bibinfo {author} {\bibfnamefont {M.}~\bibnamefont
  {Mirhosseini}}, \bibinfo {author} {\bibfnamefont {O.~S.}\ \bibnamefont
  {Magana-Loaiza}}, \bibinfo {author} {\bibfnamefont {B.~T.}\ \bibnamefont
  {Gard}}, \bibinfo {author} {\bibfnamefont {R.}~\bibnamefont {Birrittella}},
  \bibinfo {author} {\bibfnamefont {B.~E.}\ \bibnamefont {Koltenbah}}, \bibinfo
  {author} {\bibfnamefont {C.~G.}\ \bibnamefont {Parazzoli}}, \bibinfo {author}
  {\bibfnamefont {B.~A.}\ \bibnamefont {Capron}}, \bibinfo {author}
  {\bibfnamefont {C.~C.}\ \bibnamefont {Gerry}}, \bibinfo {author}
  {\bibfnamefont {J.~P.}\ \bibnamefont {Dowling}},\ and\ \bibinfo {author}
  {\bibfnamefont {R.~W.}\ \bibnamefont {Boyd}},\ }\bibfield  {title} {\bibinfo
  {title} {Quantum-enhanced interferometry with weak thermal light},\ }\href
  {https://doi.org/10.1364/OPTICA.4.000487} {\bibfield  {journal} {\bibinfo
  {journal} {Optica}\ }\textbf {\bibinfo {volume} {4}},\ \bibinfo {pages} {487}
  (\bibinfo {year} {2017})}\BibitemShut {NoStop}%
\bibitem [{\citenamefont {Hong}\ \emph
  {et~al.}(2024{\natexlab{a}})\citenamefont {Hong}, \citenamefont {Riley~B.},
  \citenamefont {Bertoni}, \citenamefont {You},\ and\ \citenamefont
  {Magaña-Loaiza}}]{Hong_Mingyuan}%
  \BibitemOpen
  \bibfield  {author} {\bibinfo {author} {\bibfnamefont {M.}~\bibnamefont
  {Hong}}, \bibinfo {author} {\bibfnamefont {D.}~\bibnamefont {Riley~B.}},
  \bibinfo {author} {\bibfnamefont {B.}~\bibnamefont {Bertoni}}, \bibinfo
  {author} {\bibfnamefont {C.}~\bibnamefont {You}},\ and\ \bibinfo {author}
  {\bibfnamefont {O.~S.}\ \bibnamefont {Magaña-Loaiza}},\ }\bibfield  {title}
  {\bibinfo {title} {Nonclassical near-field dynamics of surface plasmons},\
  }\bibfield  {journal} {\bibinfo  {journal} {Nat. Phys.}\ }\textbf {\bibinfo
  {volume} {12}},\ \href {https://doi.org/10.1038/s41567-024-02426-y}
  {10.1038/s41567-024-02426-y} (\bibinfo {year}
  {2024}{\natexlab{a}})\BibitemShut {NoStop}%
\bibitem [{\citenamefont {Gerry}\ \emph {et~al.}(2005)\citenamefont {Gerry},
  \citenamefont {Knight},\ and\ \citenamefont
  {Knight}}]{gerry2005introductory}%
  \BibitemOpen
  \bibfield  {author} {\bibinfo {author} {\bibfnamefont {C.}~\bibnamefont
  {Gerry}}, \bibinfo {author} {\bibfnamefont {P.}~\bibnamefont {Knight}},\ and\
  \bibinfo {author} {\bibfnamefont {P.~L.}\ \bibnamefont {Knight}},\
  }\href@noop {} {\emph {\bibinfo {title} {Introductory quantum optics}}}\
  (\bibinfo  {publisher} {Cambridge university press},\ \bibinfo {year}
  {2005})\BibitemShut {NoStop}%
\bibitem [{\citenamefont {Olsen}\ \emph {et~al.}(2002)\citenamefont {Olsen},
  \citenamefont {Plimak},\ and\ \citenamefont {Khoury}}]{OLSEN2002373}%
  \BibitemOpen
  \bibfield  {author} {\bibinfo {author} {\bibfnamefont {M.}~\bibnamefont
  {Olsen}}, \bibinfo {author} {\bibfnamefont {L.}~\bibnamefont {Plimak}},\ and\
  \bibinfo {author} {\bibfnamefont {A.}~\bibnamefont {Khoury}},\ }\bibfield
  {title} {\bibinfo {title} {Dynamical quantum statistical effects in optical
  parametric processes},\ }\href
  {https://doi.org/https://doi.org/10.1016/S0030-4018(01)01711-4} {\bibfield
  {journal} {\bibinfo  {journal} {Opt. Commun.}\ }\textbf {\bibinfo {volume}
  {201}},\ \bibinfo {pages} {373\textendash380} (\bibinfo {year}
  {2002})}\BibitemShut {NoStop}%
\bibitem [{\citenamefont {Kondakci}\ \emph {et~al.}(2015)\citenamefont
  {Kondakci}, \citenamefont {Abouraddy},\ and\ \citenamefont
  {Saleh}}]{kondakci2015}%
  \BibitemOpen
  \bibfield  {author} {\bibinfo {author} {\bibfnamefont {H.~E.}\ \bibnamefont
  {Kondakci}}, \bibinfo {author} {\bibfnamefont {A.~F.}\ \bibnamefont
  {Abouraddy}},\ and\ \bibinfo {author} {\bibfnamefont {B.~E.~A.}\ \bibnamefont
  {Saleh}},\ }\bibfield  {title} {\bibinfo {title} {A photonic thermalization
  gap in disordered lattices},\ }\href {https://doi.org/10.1038/nphys3482}
  {\bibfield  {journal} {\bibinfo  {journal} {Nat. Phys.}\ }\textbf {\bibinfo
  {volume} {11}},\ \bibinfo {pages} {930} (\bibinfo {year} {2015})}\BibitemShut
  {NoStop}%
\bibitem [{\citenamefont {Ou}(2007)}]{ou2007multi}%
  \BibitemOpen
  \bibfield  {author} {\bibinfo {author} {\bibfnamefont {Z.-Y.~J.}\
  \bibnamefont {Ou}},\ }\href@noop {} {\emph {\bibinfo {title} {Multi-photon
  quantum interference}}},\ Vol.~\bibinfo {volume} {43}\ (\bibinfo  {publisher}
  {Springer},\ \bibinfo {year} {2007})\BibitemShut {NoStop}%
\bibitem [{\citenamefont {You}\ and\ \citenamefont
  {et~al.}(2021{\natexlab{b}})}]{you2021scalable}%
  \BibitemOpen
  \bibfield  {author} {\bibinfo {author} {\bibfnamefont {C.}~\bibnamefont
  {You}}\ and\ \bibinfo {author} {\bibnamefont {et~al.}},\ }\bibfield  {title}
  {\bibinfo {title} {Scalable multiphoton quantum metrology with neither pre-
  nor post-selected measurements},\ }\href {https://doi.org/10.1063/5.0063294}
  {\bibfield  {journal} {\bibinfo  {journal} {Appl. Phys. Rev.}\ }\textbf
  {\bibinfo {volume} {8}},\ \bibinfo {pages} {041406} (\bibinfo {year}
  {2021}{\natexlab{b}})}\BibitemShut {NoStop}%
\bibitem [{\citenamefont {Goodman}(2008)}]{goodman2008introduction}%
  \BibitemOpen
  \bibfield  {author} {\bibinfo {author} {\bibfnamefont {J.~W.}\ \bibnamefont
  {Goodman}},\ }\href@noop {} {\emph {\bibinfo {title} {Introduction to Fourier
  optics}}}\ (\bibinfo {year} {2008})\BibitemShut {NoStop}%
\bibitem [{\citenamefont {Hong}\ \emph {et~al.}(2023)\citenamefont {Hong},
  \citenamefont {Miller}, \citenamefont {León-Montiel}, \citenamefont {You},\
  and\ \citenamefont {Magaña-Loaiza}}]{Hong2023}%
  \BibitemOpen
  \bibfield  {author} {\bibinfo {author} {\bibfnamefont {M.}~\bibnamefont
  {Hong}}, \bibinfo {author} {\bibfnamefont {A.}~\bibnamefont {Miller}},
  \bibinfo {author} {\bibfnamefont {R.~d.~J.}\ \bibnamefont {León-Montiel}},
  \bibinfo {author} {\bibfnamefont {C.}~\bibnamefont {You}},\ and\ \bibinfo
  {author} {\bibfnamefont {O.~S.}\ \bibnamefont {Magaña-Loaiza}},\ }\bibfield
  {title} {\bibinfo {title} {Engineering super-poissonian photon statistics of
  spatial light modes},\ }\href
  {https://doi.org/https://doi.org/10.1002/lpor.202300117} {\bibfield
  {journal} {\bibinfo  {journal} {Laser Photonics Rev.}\ }\textbf {\bibinfo
  {volume} {17}},\ \bibinfo {pages} {2300117} (\bibinfo {year}
  {2023})}\BibitemShut {NoStop}%
\bibitem [{\citenamefont {Agarwal}(2012)}]{agarwal2012quantum}%
  \BibitemOpen
  \bibfield  {author} {\bibinfo {author} {\bibfnamefont {G.~S.}\ \bibnamefont
  {Agarwal}},\ }\href@noop {} {\emph {\bibinfo {title} {Quantum optics}}}\
  (\bibinfo  {publisher} {Cambridge University Press},\ \bibinfo {year}
  {2012})\BibitemShut {NoStop}%
\bibitem [{\citenamefont {Mandel}(1979)}]{Mandel:79}%
  \BibitemOpen
  \bibfield  {author} {\bibinfo {author} {\bibfnamefont {L.}~\bibnamefont
  {Mandel}},\ }\bibfield  {title} {\bibinfo {title} {Sub-poissonian photon
  statistics in resonance fluorescence},\ }\href
  {https://doi.org/10.1364/OL.4.000205} {\bibfield  {journal} {\bibinfo
  {journal} {Opt. Lett.}\ }\textbf {\bibinfo {volume} {4}},\ \bibinfo {pages}
  {205\textendash207} (\bibinfo {year} {1979})}\BibitemShut {NoStop}%
\bibitem [{\citenamefont {Lawrie}\ \emph {et~al.}(2019)\citenamefont {Lawrie},
  \citenamefont {Lett}, \citenamefont {Marino},\ and\ \citenamefont
  {Pooser}}]{Lawrie2019SqueezedLight}%
  \BibitemOpen
  \bibfield  {author} {\bibinfo {author} {\bibfnamefont {B.~J.}\ \bibnamefont
  {Lawrie}}, \bibinfo {author} {\bibfnamefont {P.~D.}\ \bibnamefont {Lett}},
  \bibinfo {author} {\bibfnamefont {A.~M.}\ \bibnamefont {Marino}},\ and\
  \bibinfo {author} {\bibfnamefont {R.~C.}\ \bibnamefont {Pooser}},\ }\bibfield
   {title} {\bibinfo {title} {Quantum sensing with squeezed light},\ }\href
  {https://doi.org/10.1021/acsphotonics.9b00250} {\bibfield  {journal}
  {\bibinfo  {journal} {ACS Photonics}\ }\textbf {\bibinfo {volume} {6}},\
  \bibinfo {pages} {1307} (\bibinfo {year} {2019})}\BibitemShut {NoStop}%
\bibitem [{\citenamefont {Dawkins}\ \emph {et~al.}(2024)\citenamefont
  {Dawkins}, \citenamefont {Hong}, \citenamefont {You},\ and\ \citenamefont
  {Magana-Loaiza}}]{dawkins2024quantum}%
  \BibitemOpen
  \bibfield  {author} {\bibinfo {author} {\bibfnamefont {R.~B.}\ \bibnamefont
  {Dawkins}}, \bibinfo {author} {\bibfnamefont {M.}~\bibnamefont {Hong}},
  \bibinfo {author} {\bibfnamefont {C.}~\bibnamefont {You}},\ and\ \bibinfo
  {author} {\bibfnamefont {O.~S.}\ \bibnamefont {Magana-Loaiza}},\ }\bibfield
  {title} {\bibinfo {title} {The quantum gaussian-schell model: A link between
  classical and quantum optics},\ }\href@noop {} {\bibfield  {journal}
  {\bibinfo  {journal} {arXiv preprint arXiv:2403.09868}\ } (\bibinfo {year}
  {2024})}\BibitemShut {NoStop}%
\bibitem [{\citenamefont {Mirhosseini}\ \emph {et~al.}(2016)\citenamefont
  {Mirhosseini}, \citenamefont {Viza}, \citenamefont {Maga\~na Loaiza},
  \citenamefont {Malik}, \citenamefont {Howell},\ and\ \citenamefont
  {Boyd}}]{Mirhosseini2016PRA}%
  \BibitemOpen
  \bibfield  {author} {\bibinfo {author} {\bibfnamefont {M.}~\bibnamefont
  {Mirhosseini}}, \bibinfo {author} {\bibfnamefont {G.~I.}\ \bibnamefont
  {Viza}}, \bibinfo {author} {\bibfnamefont {O.~S.}\ \bibnamefont {Maga\~na
  Loaiza}}, \bibinfo {author} {\bibfnamefont {M.}~\bibnamefont {Malik}},
  \bibinfo {author} {\bibfnamefont {J.~C.}\ \bibnamefont {Howell}},\ and\
  \bibinfo {author} {\bibfnamefont {R.~W.}\ \bibnamefont {Boyd}},\ }\bibfield
  {title} {\bibinfo {title} {Weak-value amplification of the fast-light effect
  in rubidium vapor},\ }\href {https://doi.org/10.1103/PhysRevA.93.053836}
  {\bibfield  {journal} {\bibinfo  {journal} {Phys. Rev. A}\ }\textbf {\bibinfo
  {volume} {93}},\ \bibinfo {pages} {053836} (\bibinfo {year}
  {2016})}\BibitemShut {NoStop}%
\bibitem [{\citenamefont {Hong}\ \emph
  {et~al.}(2024{\natexlab{b}})\citenamefont {Hong}, \citenamefont {Dawkins},
  \citenamefont {Bertoni}, \citenamefont {You},\ and\ \citenamefont
  {Maga{\~n}a-Loaiza}}]{Hong2024NP}%
  \BibitemOpen
  \bibfield  {author} {\bibinfo {author} {\bibfnamefont {M.}~\bibnamefont
  {Hong}}, \bibinfo {author} {\bibfnamefont {R.~B.}\ \bibnamefont {Dawkins}},
  \bibinfo {author} {\bibfnamefont {B.}~\bibnamefont {Bertoni}}, \bibinfo
  {author} {\bibfnamefont {C.}~\bibnamefont {You}},\ and\ \bibinfo {author}
  {\bibfnamefont {O.~S.}\ \bibnamefont {Maga{\~n}a-Loaiza}},\ }\bibfield
  {title} {\bibinfo {title} {Nonclassical near-field dynamics of surface
  plasmons},\ }\bibfield  {journal} {\bibinfo  {journal} {Nat. Phys.}\ }\href
  {https://doi.org/10.1038/s41567-024-02426-y} {10.1038/s41567-024-02426-y}
  (\bibinfo {year} {2024}{\natexlab{b}})\BibitemShut {NoStop}%
\bibitem [{\citenamefont {Wen}\ \emph {et~al.}(2013)\citenamefont {Wen},
  \citenamefont {Zhang},\ and\ \citenamefont {Xiao}}]{Wen:13}%
  \BibitemOpen
  \bibfield  {author} {\bibinfo {author} {\bibfnamefont {J.}~\bibnamefont
  {Wen}}, \bibinfo {author} {\bibfnamefont {Y.}~\bibnamefont {Zhang}},\ and\
  \bibinfo {author} {\bibfnamefont {M.}~\bibnamefont {Xiao}},\ }\bibfield
  {title} {\bibinfo {title} {The talbot effect: recent advances in classical
  optics, nonlinear optics, and quantum optics},\ }\href
  {https://doi.org/10.1364/AOP.5.000083} {\bibfield  {journal} {\bibinfo
  {journal} {Adv. Opt. Photon.}\ }\textbf {\bibinfo {volume} {5}},\ \bibinfo
  {pages} {83\textendash130} (\bibinfo {year} {2013})}\BibitemShut {NoStop}%
\bibitem [{\citenamefont {Batarseh}\ and\ \citenamefont
  {et~al.}(2018)}]{Batarseh:18}%
  \BibitemOpen
  \bibfield  {author} {\bibinfo {author} {\bibfnamefont {M.}~\bibnamefont
  {Batarseh}}\ and\ \bibinfo {author} {\bibnamefont {et~al.}},\ }\bibfield
  {title} {\bibinfo {title} {Passive sensing around the corner using spatial
  coherence},\ }\href {https://doi.org/10.1038/s41467-018-05985-w} {\bibfield
  {journal} {\bibinfo  {journal} {Nat. Commun.}\ }\textbf {\bibinfo {volume}
  {9}},\ \bibinfo {pages} {3629} (\bibinfo {year} {2018})}\BibitemShut
  {NoStop}%
\bibitem [{\citenamefont {Combescure}\ \emph {et~al.}(2012)\citenamefont
  {Combescure}, \citenamefont {Robert}, \citenamefont {Combescure},\ and\
  \citenamefont {Robert}}]{combescure2012quadratic}%
  \BibitemOpen
  \bibfield  {author} {\bibinfo {author} {\bibfnamefont {M.}~\bibnamefont
  {Combescure}}, \bibinfo {author} {\bibfnamefont {D.}~\bibnamefont {Robert}},
  \bibinfo {author} {\bibfnamefont {M.}~\bibnamefont {Combescure}},\ and\
  \bibinfo {author} {\bibfnamefont {D.}~\bibnamefont {Robert}},\ }\bibfield
  {title} {\bibinfo {title} {The quadratic hamiltonians},\ }\href@noop {}
  {\bibfield  {journal} {\bibinfo  {journal} {Coherent States and Applications
  in Mathematical Physics}\ ,\ \bibinfo {pages} {59}} (\bibinfo {year}
  {2012})}\BibitemShut {NoStop}%
\bibitem [{\citenamefont {Saleh}\ and\ \citenamefont
  {Teich}(2019)}]{saleh2019fundamentals}%
  \BibitemOpen
  \bibfield  {author} {\bibinfo {author} {\bibfnamefont {B.~E.}\ \bibnamefont
  {Saleh}}\ and\ \bibinfo {author} {\bibfnamefont {M.~C.}\ \bibnamefont
  {Teich}},\ }\href@noop {} {\emph {\bibinfo {title} {Fundamentals of
  photonics}}}\ (\bibinfo  {publisher} {john Wiley \& sons},\ \bibinfo {year}
  {2019})\BibitemShut {NoStop}%
\end{thebibliography}%

\clearpage

\onecolumngrid

\section{Unpolarized Multimode Thermal Light}

In our experiment, we utilized unpolarized multimode thermal light. The light is thermal such that the electric field $E^{(+)}(x)$ obeys complex-Gaussian statistics at each point $x$, and that the mean $\braket{E^{(+)}(x)}$ is zero. It is unpolarized such that it is an equal mixture between two orthogonal polarizations (here we choose horizontal ($H$) and vertical ($V$)). Finally, any two spatial projections on this source will be statistically independent. It is our goal to study the quantum properties of such a source as it propagates through our experimental setup. We will now present a sufficient quantum description for such a light source before propagation. 

The generation of unpolarized multimode thermal light is accomplished by mixing coherent states with different amplitudes \cite{ou2007multi}. We then pixelize the source and assume that each pixel obeys independent polarization statistics. Such a source can then be written as
\begin{equation}
    \hat{\rho} = \int d\Sigma \bigotimes_{\boldsymbol{s}} \Big(|\alpha\rangle\langle\alpha|_{\Sigma,H,\boldsymbol{s}}+|\alpha\rangle\langle\alpha|_{\Sigma,V,\boldsymbol{s}}\Big).
\end{equation}
Here, each $\boldsymbol{s}$ represents the position of a pixel, $\alpha$ is a coherent amplitude, and the coherent states $|\alpha\rangle_{\Sigma,B,\boldsymbol{s}}$ are defined by the modes
\begin{equation}
    \hat{a}_{\Sigma,B,\boldsymbol{s}} = \int d\boldsymbol{\rho}\text{ Rect}\left[(\boldsymbol{s}-\boldsymbol{\rho})/d\right] \Sigma(\boldsymbol{\rho})\hat{a}_{B}(\boldsymbol{\rho}),
\end{equation}
where $d$ is the side-length of each pixel.
In the integral, $\Sigma(\boldsymbol{\rho})$ represents one instance of a random complex electric field profile. Writing $\Sigma(\boldsymbol{\rho}_i) \equiv \Sigma_i$, the action of this functional integral is characterized by the formula
\begin{equation}
    \int d\Sigma\text{ } f(\Sigma_1,...,\Sigma_n) = \int d^2\Sigma_1...d^2\Sigma_n\text{ } \frac{1}{(2\pi)^n\sqrt{|\boldsymbol{\Gamma}|}}e^{-\frac{1}{2}(\boldsymbol{r}-\boldsymbol{\mu})^T \boldsymbol{\Gamma}^{-1}(\boldsymbol{r}-\boldsymbol{\mu})}f(\Sigma_1,...,\Sigma_n),
\end{equation}
where $\boldsymbol{r} \equiv \Big(\text{Re}[\Sigma_1],\text{Im}[\Sigma_1],...,\text{Re}[\Sigma_n],\text{Im}[\Sigma_n]\Big)$, $\boldsymbol{\mu} = \braket{\boldsymbol{r}}$, $\boldsymbol{\Gamma}$ is the covariance matrix of $\boldsymbol{r}$, and $|\cdot|$ represents the determinant operation. Note that, in the case of thermal statistics, $\boldsymbol{\mu} = \boldsymbol{0}$. To complete our description of unpolarized multimode thermal light, we now determine the covariance matrix $\boldsymbol{\Gamma}$. It is easy to see that $\boldsymbol{\Gamma}$ is completely determined by $\braket{\Sigma(\boldsymbol{\rho}_1)\Sigma(\boldsymbol{\rho}_2)}$ and $\braket{\Sigma^*(\boldsymbol{\rho}_1)\Sigma(\boldsymbol{\rho}_2)}$. The term $\braket{\Sigma(\boldsymbol{\rho}_1)\Sigma(\boldsymbol{\rho}_2)}$ will always be $0$ in the case of thermal light, and $\braket{\Sigma^*(\boldsymbol{\rho}_1)\Sigma(\boldsymbol{\rho}_2)}$ will have the form
\begin{equation}
    \braket{\Sigma^*(\boldsymbol{\rho}_1)\Sigma(\boldsymbol{\rho}_2)} = \sqrt{\bar{n}(\boldsymbol{\rho}_1)\bar{n}(\boldsymbol{\rho}_2)} \left[\frac{1}{\pi \sigma}e^{-\frac{|\boldsymbol{\rho}_1-\boldsymbol{\rho}_2|^2}{\sigma}}\right],
\end{equation}
where $\sigma$ is assumed to be small so that $\frac{1}{\pi \sigma}e^{-\frac{|\boldsymbol{\rho}_1-\boldsymbol{\rho}_2|^2}{\sigma}} \approx \delta(\boldsymbol{\rho}_1-\boldsymbol{\rho}_2)$. Normalization gives us the mean photon number at position $\boldsymbol{\rho}$ as $\bar{n}(\boldsymbol{\rho}) = \pi\sigma/d^2$.

\section{Propagation to the Far Field}

The temporal evolution of a photon's spatial probability distribution obeys classical physics. This behavior is a direct consequence of the free-space Hamiltonian for the electromagnetic field being quadratic in its quadrature variables \cite{combescure2012quadratic,agarwal2012quantum}.
Assuming a paraxial light source, we can utilize the Fresnel diffraction formula to determine the propagated mode-structure of the unpolarized multimode thermal light \cite{saleh2019fundamentals}. The Fresnel kernel at propagation distance $z$ is given by \cite{goodman2008introduction}
\begin{equation}
    K(\boldsymbol{r},\boldsymbol{\rho},z) = \frac{e^{i k z}}{i\lambda z} e^{\frac{ik}{2z}(\boldsymbol{r}-\boldsymbol{\rho})},
\end{equation}
where $\lambda$ is the wavelength of the light source, $k = \frac{2\pi}{\lambda}$, and $\boldsymbol{r},\boldsymbol{\rho}$ represent positions in the measurement plane and source plane respectively. We can calculate the propagated mode structure as
\begin{equation}
    E^{(+)}(\boldsymbol{r},z) = \int d\boldsymbol{\rho} K(\boldsymbol{r},\boldsymbol{\rho},z)E^{(+)}(\boldsymbol{r},0).
\end{equation}
Given a mode $\hat{a}_0 = \int d\boldsymbol{r}  \text{ } f(\boldsymbol{r}) \hat{a}_i(\boldsymbol{r})$ with arbitrary polarization $i$, the resulting mode in the far-field is given by
\begin{equation}
    \hat{a}_z = \int d\boldsymbol{r}\left[\int d\boldsymbol{\rho}\text{ }K^*(\boldsymbol{r},\boldsymbol{\rho},z)f(\boldsymbol{\rho})\right]\hat{a}_i(\boldsymbol{r}).
\end{equation}
Importantly, we note that the operator-valued distribution $\hat{a}_i(\boldsymbol{r})$ is not integrated against the Fresnel kernel.

\section{Computing the Correlation Matrix}

In our experiment, a linear polarization grating with a position-dependent polarization angle filters the multimode unpolarized light, and the beam is propagated in free-space \cite{you2023multiphoton}. In the far-field, we make measurements with two point-detectors which are able to post-select on a particular configuration of polarizations. To be explicit, the operator representing a measurement of the first-order correlation at position $\boldsymbol{\rho}$ is given by $\hat{a}^\dagger_i (\boldsymbol{\rho})\hat{a}_j(\boldsymbol{\rho})$ where $i,j\in \{H,V\}$. We represent the transformation of the polarization grating with the matrix
\begin{equation}
    \boldsymbol{P}(x) = \begin{pmatrix}
        \cos^2\left(\frac{\pi x}{L}\right)& \cos \left(\frac{\pi x}{L}\right) \sin \left(\frac{\pi x}{L}\right)\\
        \cos \left(\frac{\pi x}{L}\right)\sin \left(\frac{\pi x}{L}\right)& \sin^2\left(\frac{\pi x}{L}\right)\\
        \sin \left(\frac{\pi x}{L}\right)&\cos \left(\frac{\pi x}{L}\right)
    \end{pmatrix},
\end{equation}
where $L$ is the width of the polarization grating and the represented transformation is given by
\begin{equation}
    \hat{a}_B(\boldsymbol{\rho}) = P_{HB}(\rho_x)\hat{a}_H(\boldsymbol{\rho}) + P_{VB}(\rho_x)\hat{a}_V(\boldsymbol{\rho}) + P_{\emptyset B}(\rho_x)\hat{a}_\emptyset(\boldsymbol{\rho}),
\end{equation}
for $B\in \{H,V\}$ and $\boldsymbol{\rho} = \rho_x\hat{\boldsymbol{x}} + \rho_y\hat{\boldsymbol{y}}$. The $\hat{a}_\emptyset(\boldsymbol{\rho})$ mode represents photon-loss at the polarization grating. Therefore, immediately after the linear polarizer, the mode structure is given by
\begin{equation}
    \hat{a}_{\Sigma,B,\boldsymbol{s}} = \int d\boldsymbol{\rho}\text{ Rect}\left[(\boldsymbol{s}-\boldsymbol{\rho})/d\right]\Sigma(\boldsymbol{\rho}) \Big[P_{HB}(\rho_x)\hat{a}_H(\boldsymbol{\rho}) + P_{VB}(\rho_x)\hat{a}_V(\boldsymbol{\rho}) + P_{\emptyset B}(\rho_x)\hat{a}_\emptyset(\boldsymbol{\rho})\Big].
\end{equation}
Consequently, in the far-field, it is given by
\begin{equation}
    \hat{a}_{\Sigma,B,\boldsymbol{s},z} = \int d\boldsymbol{r} \text{ }\left(\int d\boldsymbol{\rho} K^*(\boldsymbol{r},\boldsymbol{\rho},z)\text{ Rect}\left[(\boldsymbol{s}-\boldsymbol{\rho})/d\right]\Sigma(\boldsymbol{\rho})\Big[P_{HB}(\rho_x)\hat{a}_H(\boldsymbol{r}) + P_{VB}(\rho_x)\hat{a}_V(\boldsymbol{r}) + P_{\emptyset B}(\rho_x)\hat{a}_\emptyset(\boldsymbol{r})\Big]\right).
\end{equation}
We now make a couple of approximations to simplify our calculations. First, we suppose that $\text{ Rect}\left[(\boldsymbol{s}-\boldsymbol{\rho})/d\right]\approx \text{ Rect}\left[\boldsymbol{\rho}/L\right]$ for all pixel positions $\boldsymbol{s}$. Then, we assume that light at each position in the far-field had originated primarily from a single pixel. With these approximations, we are able to write the propagated state in the form
\begin{equation}
    \hat{\rho}_{z} = \int d\Sigma \bigotimes_{\boldsymbol{r}} \Big(|\alpha\rangle\langle\alpha|_{\Sigma,H,\boldsymbol{r},z}+|\alpha\rangle\langle\alpha|_{\Sigma,V,\boldsymbol{r},z}\Big),
\end{equation}
where the mode structure for each $\boldsymbol{r}$ is now given by
\begin{equation}
\begin{aligned}
    \hat{a}_{\Sigma,B,\boldsymbol{r},z} \approx \int d\boldsymbol{r}' & \text{ Rect}\left[\frac{(\boldsymbol{r}-\boldsymbol{r}')}{d'}\right]\Bigg(\int d\boldsymbol{\rho} K^*(\boldsymbol{r}',\boldsymbol{\rho},z)\text{ Rect}\left[\frac{\boldsymbol{\rho}}{d}\right]\Sigma(\boldsymbol{\rho})\\
    &\times\Big[P_{HB}(\rho_x)\hat{a}_H(\boldsymbol{r}) + P_{VB}(\rho_x)\hat{a}_V(\boldsymbol{r}) + P_{\emptyset B}(\rho_x)\hat{a}_\emptyset(\boldsymbol{r})\Big]\Bigg),
\end{aligned} 
\end{equation}
where $d'$ is the width of each pixel in the measurement plane. From here, we can compute the second-order correlation functions for various polarization projections as 
\begin{equation}
    \begin{aligned}
        G^{(2)}_{ijkl}(\boldsymbol{r}_1,\boldsymbol{r}_2,z) &= \int d\Sigma \text{ }\sum_{A,B}\Big(\bra{\alpha}_{\Sigma, A,\boldsymbol{r}_1, z}\bra{\alpha}_{\Sigma, B,\boldsymbol{r}_2, z} \Big)\hat{a}^\dagger_i(\boldsymbol{r}_1)\hat{a}^\dagger_j(\boldsymbol{r}_2)\hat{a}_k(\boldsymbol{r}_1)\hat{a}_l(\boldsymbol{r}_2)\Big(\ket{\alpha}_{\Sigma, A,\boldsymbol{r}_1, z}\ket{\alpha}_{\Sigma, B,\boldsymbol{r}_2, z} \Big)\\
        &= \int d\Sigma \text{ } |\alpha|^4 \int d\boldsymbol{\rho}_1 d\boldsymbol{\rho}_2 d\boldsymbol{\rho}_3 d\boldsymbol{\rho}_4 K^*(\boldsymbol{r}_1,\boldsymbol{\rho}_1,z)K^*(\boldsymbol{r}_2,\boldsymbol{\rho}_2,z)\\
        &\text{ }\text{ }\text{ }\times K(\boldsymbol{r}_1,\boldsymbol{\rho}_3,z)K(\boldsymbol{r}_2,\boldsymbol{\rho}_4,z)\Sigma(\boldsymbol{\rho}_1)\Sigma(\boldsymbol{\rho}_2)\Sigma^*(\boldsymbol{\rho}_3)\Sigma^*(\boldsymbol{\rho}_4)\\
        &\text{ }\text{ }\text{ }\times \sum_{A,B}P_{iA}(\rho_{1x})P_{jB}(\rho_{2x})P_{kA}(\rho_{3x})P_{lB}(\rho_{4x})\\
        &\approx |\alpha|^4\int d\boldsymbol{\rho}_1 d\boldsymbol{\rho}_2 d\boldsymbol{\rho}_3 d\boldsymbol{\rho}_4 K^*(\boldsymbol{r}_1,\boldsymbol{\rho}_1,z)K^*(\boldsymbol{r}_2,\boldsymbol{\rho}_2,z)K(\boldsymbol{r}_1,\boldsymbol{\rho}_3,z)K(\boldsymbol{r}_2,\boldsymbol{\rho}_4,z)\\
        &\text{ }\text{ }\text{ }\times \frac{\pi^2\sigma^2}{L^4}\text{Rect}(\frac{\boldsymbol{\rho}_1}{L})\text{Rect}(\frac{\boldsymbol{\rho}_2}{L})\text{Rect}(\frac{\boldsymbol{\rho}_3}{L})\text{Rect}(\frac{\boldsymbol{\rho}_4}{L})\\
        &\text{ }\text{ }\text{ }\times\Big[\delta(\boldsymbol{\rho}_1-\boldsymbol{\rho}_3)\delta(\boldsymbol{\rho}_2-\boldsymbol{\rho}_4) + \delta(\boldsymbol{\rho}_1-\boldsymbol{\rho}_4)\delta(\boldsymbol{\rho}_2-\boldsymbol{\rho}_3)\Big]\\
        &\text{ }\text{ }\text{ }\times \frac{1}{4}\Big[P_{iH}(\rho_{1x})P_{jH}(\rho_{2x})P_{kH}(\rho_{3x})P_{lH}(\rho_{4x}) + P_{iH}(\rho_{1x})P_{jH}(\rho_{2x})P_{kV}(\rho_{3x})P_{lV}(\rho_{4x})\\
        &\text{ }\text{ }\text{ } + P_{iV}(\rho_{1x})P_{jV}(\rho_{2x})P_{kH}(\rho_{3x})P_{lH}(\rho_{4x}) + P_{iV}(\rho_{1x})P_{jV}(\rho_{2x})P_{kV}(\rho_{3x})P_{lV}(\rho_{4x})\Big]\\
        &\equiv I_0 \int d\boldsymbol{\rho}_1 d\boldsymbol{\rho}_2 d\boldsymbol{\rho}_3 d\boldsymbol{\rho}_4 \text{ } F(\boldsymbol{r}_1,\boldsymbol{r}_2,\boldsymbol{\rho}_1,\boldsymbol{\rho}_2,\boldsymbol{\rho}_3,\boldsymbol{\rho}_4,z) \\
        &\text{ }\text{ }\text{ }\times\Big[\delta(\boldsymbol{\rho}_1-\boldsymbol{\rho}_3)\delta(\boldsymbol{\rho}_2-\boldsymbol{\rho}_4) + \delta(\boldsymbol{\rho}_1-\boldsymbol{\rho}_4)\delta(\boldsymbol{\rho}_2-\boldsymbol{\rho}_3)\Big],
    \end{aligned}
\end{equation}
where we have defined $I_0 = \pi^2 \sigma^2 |\alpha|^4/L^4$ and
\begin{equation}
    \begin{aligned}
    F(\boldsymbol{r}_1,\boldsymbol{r}_2,\boldsymbol{\rho}_1,\boldsymbol{\rho}_2,\boldsymbol{\rho}_3,\boldsymbol{\rho}_4,z)  =&\text{ } K^*(\boldsymbol{r}_1,\boldsymbol{\rho}_1,z)K^*(\boldsymbol{r}_2,\boldsymbol{\rho}_2,z)K(\boldsymbol{r}_1,\boldsymbol{\rho}_3,z)K(\boldsymbol{r}_2,\boldsymbol{\rho}_4,z)\\
        &\times \text{Rect}(\frac{\boldsymbol{\rho}_1}{L})\text{Rect}(\frac{\boldsymbol{\rho}_2}{L})\text{Rect}(\frac{\boldsymbol{\rho}_3}{L})\text{Rect}(\frac{\boldsymbol{\rho}_4}{L})\\
        &\times \frac{1}{4}\Big[P_{iH}(\rho_{1x})P_{jH}(\rho_{2x})P_{kH}(\rho_{3x})P_{lH}(\rho_{4x}) + P_{iH}(\rho_{1x})P_{jH}(\rho_{2x})P_{kV}(\rho_{3x})P_{lV}(\rho_{4x})\\
        &\text{ }\text{ }\text{ }\text{ }\text{ }+ P_{iV}(\rho_{1x})P_{jV}(\rho_{2x})P_{kH}(\rho_{3x})P_{lH}(\rho_{4x}) + P_{iV}(\rho_{1x})P_{jV}(\rho_{2x})P_{kV}(\rho_{3x})P_{lV}(\rho_{4x})\Big].
    \end{aligned}
\end{equation}
These definitions allow for a drastically simplified $G^{(2)}_{ijkl}(\boldsymbol{r}_1,\boldsymbol{r}_2,z)$, and they are used in the main body of the paper. From here, each $G^{(2)}_{ijkl}(\boldsymbol{r}_1,\boldsymbol{r}_2,z)$ can be calculated explicitly. Furthermore, we can use a similar approach to show that $G^{(1)}_{i,j}(\boldsymbol{r},z) = \braket{\hat{a}^\dagger_i(\boldsymbol{r})\hat{a}_j(\boldsymbol{r})} = \sqrt{I_0}L/(2 z^2 \lambda^2)$. Using these, the normalized second-order correlation functions $g^{(2)}_{ijkl}(\boldsymbol{r}_1,\boldsymbol{r}_2,z)$ can be calculated, and this list is presented in the next section. Notably, these results are in agreement with our previous theoretical approach and our experimental data \cite{you2023multiphoton}.

\section{List of Relevant Second-Order Correlation Functions}

Here we explicitly write the relevant second-order coherence functions studied in our experiment. In this section, we are using the shorthands $\sinc(\nu) \equiv \sin(\pi \nu)/(\pi \nu)$ and $\nu = L(r_{1x}-r_{2x})/(\lambda z)$.
\begin{equation}
    \begin{aligned}
        g^{(2)}_{\text{HHHH}}(\nu) =& \frac{1}{16}(10 \sinc(\nu)^2 + 2 (6\sinc(\nu + 1) + \sinc(\nu + 2) + 6\sinc(1-\nu) + \sinc(2-\nu))\sinc(\nu)\\
        & + 6\sinc(\nu+1)^2 + \sinc(\nu+2)^2 + 6\sinc(1-\nu)^2 + \sinc(2-\nu)^2 \\
        & + 4\sinc(\nu+1)\sinc(\nu+2) + 4(\sinc(\nu_1)+\sinc(2-\nu))\sinc(1-\nu)+16),\\
        g^{(2)}_{\text{HVHV}}(\nu) =& \frac{1}{16}(2\sinc(\nu)^2 - 2(\sinc(\nu+2)+\sinc(2-\nu))\sinc(\nu) + 2(\sinc(1-\nu)\\
        &- \sinc(\nu+1))^2 + \sinc(\nu+2)^2 + \sinc(2-\nu)^2 + 16),\\
        g^{(2)}_{\text{VHHV}}(\nu) =& \frac{1}{16}(6\sinc(\nu)^2 - 2(\sinc(\nu+2)+\sinc(2-\nu))\sinc(\nu) + 2(\sinc(1-\nu)\\
        &-\sinc(\nu+1))^2 - \sinc(\nu+2)^2 - \sinc(2-\nu)^2),\\
        g^{(2)}_{\text{HHVV}}(\nu) =& \frac{1}{16}(2\sinc(\nu)^2 - 2(\sinc(\nu+2)+\sinc(2-\nu))\sinc(\nu)\\
        &+ 2(\sinc(1-\nu)-\sinc(\nu+1))^2 + \sinc(\nu+2)^2 + \sinc(2-\nu)^2).
    \end{aligned}
\end{equation}

\section{Propagation of the Photon Number Distribution}

In this section, we present a method for determining the photon number distribution in different detection planes. In doing so, we can study the dynamics of multiphoton wavepackets. It will be challenging to compute the photon number distribution directly from $\hat{\rho}_z$, but we can avoid this difficulty by recognizing that
\begin{equation}
    \eta_{AB}(\boldsymbol{r}) = \int d\boldsymbol{\rho} K^*(\boldsymbol{r},\boldsymbol{\rho},z)\text{ Rect}\left[\boldsymbol{\rho}/L\right]\Sigma(\boldsymbol{\rho})P_{AB}(\rho_x)
\end{equation}
follows Gaussian statistics as a result of $\Sigma(\boldsymbol{\rho})$ obeying Gaussian statistics. Each $\eta_{AB}(\boldsymbol{r})$ represents one instance of a coherent state, and so by determining the probability distribution of the $\eta_{AB}(\boldsymbol{r})$ we can determine the effective quantum state as measured by our detectors. For post-selected polarization $ijkl$ at positions $\boldsymbol{r}_1,\boldsymbol{r_2}$, we will need the probability distribution for $\eta_{iA}(\boldsymbol{r_1}),\eta_{jB}(\boldsymbol{r_2}),\eta_{kC}^*(\boldsymbol{r_1}),\eta_{lD}^*(\boldsymbol{r_2}) \equiv \alpha_{iA},\alpha_{jB},\alpha_{kC},\alpha_{lD}$ where $A,B,C,D\in \{H,V\}$. Denoting $\boldsymbol{t} = (\text{Re}[\alpha_{iA}],\text{Im}[\alpha_{iA}],...,\text{Re}[\alpha_{lD}],\text{Im}[\alpha_{lD}])$, the desired probability distribution is given by
\begin{equation}
    P_{iAjBkClD}(\boldsymbol{t}) = \frac{1}{(2\pi)^4\sqrt{|\boldsymbol{\Gamma}|}}e^{-\frac{1}{2}(\boldsymbol{t}-\boldsymbol{\mu})^T \boldsymbol{\Gamma}^{-1}(\boldsymbol{t}-\boldsymbol{\mu})},
\end{equation}
where $\boldsymbol{\mu} = \braket{\boldsymbol{t}}$ and $\Gamma_{nm} = \braket{t_n t_m} - \braket{t_n}\braket{t_m}$. With this, the resulting state describing these statistics is now
\begin{equation}
    \hat{\rho}_{iAjBkClD}(z) = \int d^2 \alpha_{iA}d^2 \alpha_{jB}d^2 \alpha_{kC}d^2 \alpha_{lD} P_{iAjBkClD}(\alpha_{iA}, \alpha_{jB}, \alpha_{kC},\alpha_{lD}) |\alpha_{kC},\alpha_{lD}\rangle\langle \alpha_{iA}, \alpha_{jB}|.
\end{equation}
It then follows that the total state is given by
\begin{equation}
    \hat{\rho}_{ijkl}(z) = \frac{1}{4}\left[\hat{\rho}_{iHjHkHlH}(z) + \hat{\rho}_{iHjHkVlV}(z) + \hat{\rho}_{iVjVkHlH}(z) + \hat{\rho}_{iVjVkVlV}(z)\right],
\end{equation}
and thus that the photon-number distribution $p(n_1,n_2)$ can be calculated via
\begin{equation}
    p(n_1,n_2,z) = \text{Tr}\left[\hat{\rho}_{ijkl}(z) |n_1,n_2\rangle\langle n_1,n_2|\right].
\end{equation}

\section{Realization of Polarization Grating through a Spatial Light Modulator}
\begin{figure*}[!ht]
   \centering 
   \includegraphics[width=0.85\textwidth]{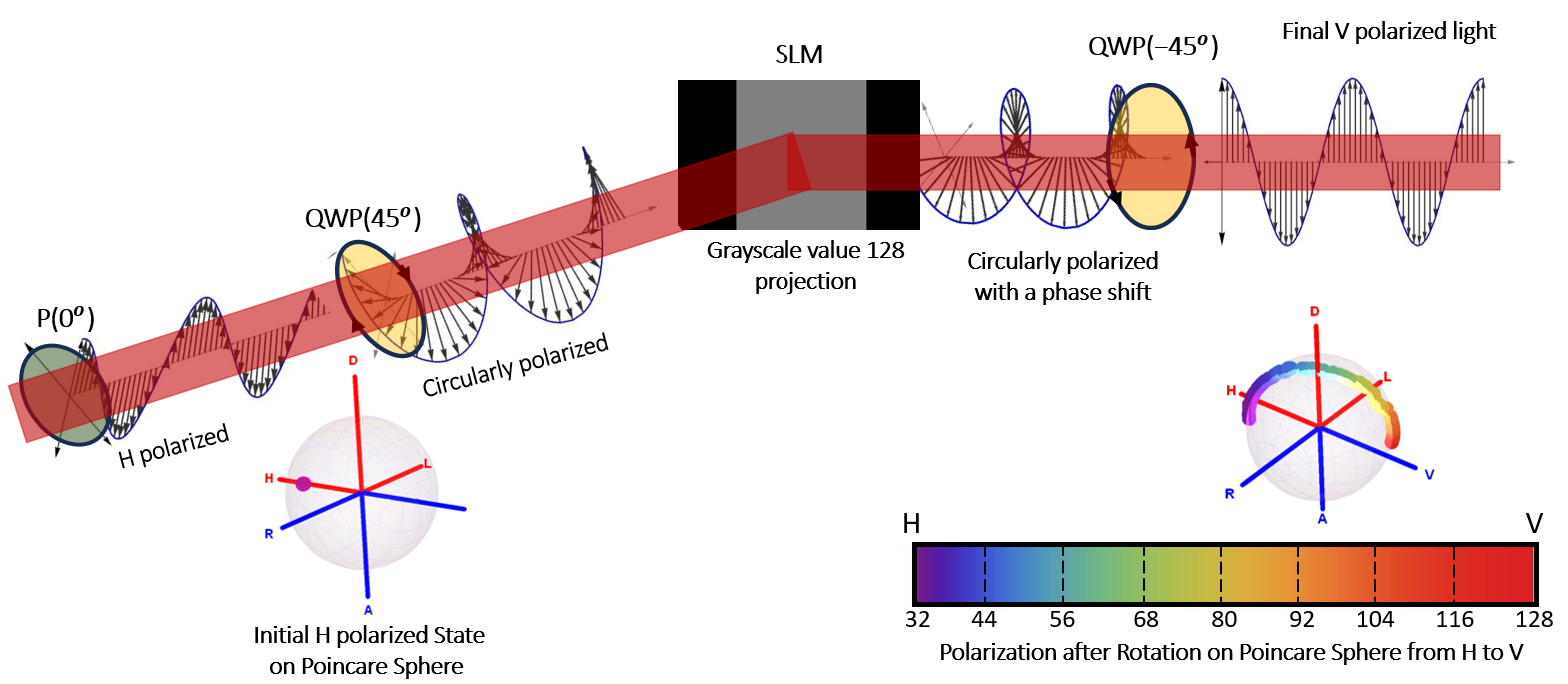}
\caption{\textbf{Schematic diagram for polarization rotation.} The illustration portrays a beam undergoing polarization rotation via polarization optics and a spatial light modulator (SLM). We characterize the polarization control ability of our experimental setup, and the corresponding results are displayed on the Poincare sphere on the bottom right.} 
\label{SI-1}
\end{figure*} 

In this section, we describe the realization of the polarization grating using polarization optics and a spatial light modulator (SLM) \cite{PRLMirhossein2014}. As shown in Fig. \ref{SI-1}, the polarization rotation of the input beam is performed using a SLM in combination with two quarter-wave plates (QWPs). Specifically, the input beam is prepared by passing it through a polarizer aligned to the H polarization. The beam first passes through a QWP at an angle of $45^\circ$. Then, this beam is imprinted on the SLM, where a gray-value image is displayed. Finally, the reflected beam passes through another QWP at an angle of $-45^\circ$. This configuration provides the ability to rotate the polarization of the incident beam in a controlled fashion. We then characterize the relationship between the polarization rotation and the gray-value displayed on the SLM. This allows us to design a gray-scale image to implement the polarization grating. By adjusting the gray-scale values across different pixels along the $x$-axis of the SLM screen, we can control the polarization at each pixel. We can thus simulate the effect of a polarization grating on an unpolarized light source.

\end{document}